\documentclass[useAMS,usenatbib,referee]{mn2e}

\usepackage{epsfig,rotate}
\usepackage{epsf}
\usepackage{graphicx}
\usepackage{natbib}
\usepackage{times}
\usepackage{float}
\newcommand{\bmth}[1]{\mbox{\boldmath${#1}$}}

\title[Dynamic tides in rotating objects ]
{Dynamic tides in rotating objects: 
a numerical investigation of inertial waves in fully convective or barotropic stars and planets}
\author[J. C. B. Papaloizou and P. B. Ivanov]{
J. C. B. Papaloizou$^{1}$\thanks{
J.C.B.Papaloizou@damtp.cam.ac.uk (JCBP), pbi20@cam.ac.uk
(PBI)}  P. B. Ivanov$^{1,2}$
\footnotemark[1]\\
$^{1}$Department of Applied Mathematics and Theoretical Physics,
University of Cambridge,\\
Centre for Mathematical Sciences, 
Wilberforce Road, Cambridge, CB3 0WA, UK \\
$^{2}$Astro Space Centre, P. N. Lebedev Physical Institute,
4/32 Profsoyuznaya Street,
Moscow, 117810, Russia}

\begin{document}

\date{Accepted Received ; in original form }

\pagerange{\pageref{firstpage}--\pageref{lastpage}} \pubyear{2002}

\maketitle

\label{firstpage}

\begin{abstract}

We perform direct numerical simulations 
of the tidal encounter of a rotating planet on  a highly eccentric 
or parabolic orbit  about a central star formulated as an initial value
problem. This approach enables us to  extend previous work
of  Ivanov \& Papaloizou to consider planet models with solid cores
and to avoid making an anelastic approximation.  

We obtain a power spectrum of the tidal response of coreless models
which enables  global inertial modes to be identified. Their frequencies
are found to be in good agreement with those obtained using either a WKBJ approach or
 the anelastic spectral approach adopted in previous work
for small planet rotation rates.
We also find that the dependence of the normal mode frequencies
on the planet angular velocity in case of higher rotation rates 
can for the most part be understood   by  applying  first
order perturbation theory to the anelastic  modes.

We calculate  the energy and angular momentum 
exchanged as a result of the tidal encounter and  for coreless models
again find good agreement with results obtained using either  the anelastic spectral method.

Models with a solid core  showed evidence of {the emission of shear layers
at critical latitudes and possibly } wave attractors
after the encounter but the total energy exchanged during the encounter
did not differ dramatically from the coreless case  { as long as the ratio of the  core radius
to the total radius was less than $50\%$,  there being hardly any difference at all when  this ratio   was  less than $25\%$
of the total radius.}  We give a physical and mathematical interpretation
of this result.

Finally we are able to validate the use of the anelastic  approximation
for both the work presented here and our previous work which led
to estimates of  circularisation rates for planets in highly eccentric orbits.

\end{abstract}

\vspace{-1cm}
\section{Introduction}

A significant number of extrasolar planets are found in close circular orbits around their central star
while others are found in highly eccentric orbits. This is an indication that planet-planet
scattering may have been important after formation and generated highly eccentric orbits that
subsequently undergo  circularisation  with attendant tidal heating ( see Papaloizou \& Terquem 2006 for a review 
and references therein).  In order to investigate such a scenario the tidal interaction of a   planet
on a highly eccentric orbit with the central star has to be
evaluated. This was considered in 
the non rotating case by Ivanov \& Papaloizou (2004)
and in the rotating coreless case using an anelastic approximation 
by Papaloizou \& Ivanov (2005) and Ivanov \& Papaloizou (2007)
(hereafter referred to as PI and IP, respectively).
The excitation of  pulsations or normal modes of the planet
is a key aspect of the tidal  problem.

In this paper we extend our previous work by relaxing the anelastic approximation
and considering planet models with solid cores.
 We investigate pulsations of  a uniformly rotating 
fully convective or barotropic object by  performing numerical simulations
of tidal encounters considered  as  initial value problems.
The formalism we adopt  does not require  an anelastic approximation  to be made and so it enables  both the response of  
the low frequency inertial  modes of oscillation as well as the higher frequency $f$
and $p$ modes to be taken into account.
As in our preceding studies  we  consider the pulsations to be  excited in an object 
referred to  hereafter  as a 'planet'  as a consequence of  moving on a parabolic orbit around a central object.

We calculate the energy and angular momentum exchanged after 
 pericentre passage for a variety of pericentre distances and  angular velocities of rotation 
of the planet. We also study the spectrum of normal modes that is excited 
and compare their properties with the analytic theory developed in
our accompanying paper Ivanov \& Papaloizou (2009) hereafter  IPN and also results obtained
using the anelastic spectral approach described in Ivanov \& Papaloizou (2007) (hereafter referred to as IP) .
This comparison finds good agreement between the different
approaches,  validating the use of the anelastic approximation for  describing the
inertial mode response. 

We also consider the effects
of introducing a solid core into the planet. Although we find evidence for the 
development of singular phenomena such as { the emission of shear layers at critical latitudes
and the formation of } wave attractors
(see Ogilvie \& Lin 2004, Ogilvie 2009 { and Rieutord \& Valdettaro 2010} and references therein)
 and  after the encounter,
the energy  exchanged is  in general not changed by  a very large amount  and  for a modest sized
core  hardly
at all. We give both a physical and mathematical explanation for this result
in appendix B.

The plan of the paper is as follows. In section 2 we review some definitions and basic equations
relegating a description of the numerical scheme we use to  solve initial value problems to appendix A.
Note that no anelastic approximation is made here.
In section 3 we go on to describe the results we obtained for the response of a simple model of a giant
planet as a polytrope of index $n=1$ to tidal encounters.  We present results for the total energy
and angular momentum transferred to a planet without a solid core for  the full range of rotation rates and  a variety
of encounter pericentre distances.  We then go on to consider planet models with solid cores
of radius $25\%$ and $50\%$ of the total radius illustrating  evidence of the 
focusing of inertial waves,  {  the emission of shear layers at critical latitudes}  and wave attractors, 
in these cases.
The role of these phenomena has  recently
been emphasised by Ogilvie \& Lin (2004) and Ogilvie (2009)
for the situation where the planet is tidally forced while  in a fixed circular orbit.
 However, in our case with the smaller core, the energy and angular momentum
transferred hardy differs from  what is found for the coreless case.
In the larger core  case the interaction is somewhat stronger with the energy transfer being about one order
of magnitude larger for the  model with a core. 

In section 4 we obtain the normal mode frequencies for the coreless planet model by  taking Fourier transforms
of the time series  provided by the post tidal interaction responses and locating peaks in the corresponding
power spectra. The normal mode frequencies are compared with those obtained using the spectral method
described in PI and also the WKBJ method described in IPN, both of which  adopted an anelastic
approximation.  The spectra obtained from the simulations
are seen to contain  $p$ and  $f$ modes in addition to the inertial modes. The  frequency locations of the main global
inertial modes are  found to be in good agreement with those obtained by other methods for azimuthal mode numbers $m=0$ and $m=2.$ 

In section  4.1.1 we investigate the dependence of the simulation
normal mode frequencies (or eigenfrequencies) on the planet
rotation rate paying a special attention to the case of inertial
waves. In the anelastic approximation it can be shown that
eigenfrequencies of inertial modes are proportional to the planet's
rotation rate, $\Omega $, so it is natural to express their values in
terms of it. This proportionality breaks down when the  anelastic
approximation is not used and the positions of the mode  eigenfrequencies
measured in units of $\Omega $  are  shifted  when it  changes.
We compare positions of the simulation normal mode frequencies with those obtained using the anelastic spectral method but with
a correction obtained by applying the first order perturbation theory given in IPN. The agreement is very good apart
for one case that is apparently affected by an avoided  crossing.

 For completeness
 we consider the tidal response of a coreless model of the type
 considered by Goodman \& Lackner(2009) that is 
 in hydrostatic equilibrium under a fixed quadratic
gravitational potential in section 5.
 These authors pointed out that  a simple analytic solution
is available and shows that no inertial mode excitation
is expected in this case. Our numerical results indicate
that  any residual  inertial mode excitation occurring
as a result of numerical effects  is very small in this case
as should be expected.
 
In section 6 we compare the total energy and angular momentum transferred to a coreless polytropic planet 
with the anelastic  spectral results already given in IP.
We conclude that these are in good agreement and therefore the anelastic approximation
is a valid approximation. Thus we do not confirm the suggestion of Goodman \& Lackner (2009) that these exchanges
could be seriously overestimated as a consequence of its use.
Finally in section 7 we discuss our results.

\section{Basic definitions and equations}
Here we adopt  a spherical polar coordinate system
$(r, \theta, \phi)$ and the associated  spherical coordinate system
 with origin at the centre of mass of the planet.

We consider the  response to  a tidal perturbation 
with associated gravitational potential $\Psi_{ext}$
written as a Fourier series  in terms of the azimuthal angle  $\phi$  in
the form
\begin{equation}
\Psi_{ext} = {\cal R}e\sum_{m=0}^{\infty}\Psi_{ext,m}\exp({im\phi}). \label{eq p1} \end{equation}
Here ${\cal R}e $ denotes that the real part is to be taken
and although the sum is in general over all $m,$ we consider 
only   $m=0$  and $2$ here. Similar Fourier decompositions
are used for the state variables of the planet. In the linear approximation
each  Fourier component responds individually to the corresponding
component of the tidal potential. 

The planet is characterised by its mass $M_{*}$, radius $R_{*}$ and
the associated frequency 
\begin{equation}
\Omega_{*}=\sqrt{GM_{*}\over R_{*}^{3}}, \label{eqn p2} 
\end{equation}
where $G$ is the gravity constant. 
The associated energy and angular momentum scales
are $E_*= GM_*^2/R_*$ and  $L_* = M_*\sqrt{GM_*R_*}$
respectively. For definiteness, we consider below a planet having approximately 
Jovian values of mass and radius: $M_{*}$=$2\times 10^{30}g$ and $R_{*}=7\times 
10^{9}cm$. Accordingly, it is assumed that $E_{*}\approx 3.81\times 10^{43}$,
$L_{*}\approx 6\times 10^{46}$ and $ \Omega_*\approx 6.23\times 10^{-4}$  in $cgs$ units. Note, however, that our results,
given in dimensional form, may be scaled to other values of $M_{*}$ and $R_{*}$ 
when expressed in the natural units introduced above.

Following our previous work (Papaloizou $\&$ Ivanov 2005, hereafter PI, Ivanov $\&$ Papaloizou 
2007, hereafter IP) we assume in this paper that the planet moves on a highly eccentric (formally,
parabolic) orbit around a source of gravity of mass $M$. In the 
parabolic limit the orbit may be characterised by its pericentre distance,
$R_{p}$. The simulations carried out in this paper started with the perturbing
mass at a distance eight times the pericentre distance. Tidal
forces being proportional to the inverse cube of the distance are negligible
beyond this point (see eg. Faber et al 2005). 
For fixed masses, as an alternative to the pericentre distance,
in this paper we define an encounter through
the  parameter
\begin{equation} \eta  =\sqrt{ M_*R_{p}^3/(M R_*^3)}, \label{eqn1} \end {equation}
first introduced by Press $\&$ Teukolsky 1977, hereafter PT. Note that
this expression for $\eta $ is strictly speaking valid only when
$M_{*}\ll M$. A more general expression is given in equation
(\ref{a2}) of appendix A.

\subsection{Perturbed equations of motion}

We assume that the planet is rotating with uniform angular velocity
$\bmth {\Omega }$ directed perpendicular to the orbital plane. 
The hydrodynamic equations for the perturbed quantities 
take the simplest form in the rotating
frame which we use later on.

Since the planet is fully convective,
the entropy per unit of mass of the planetary gas remains
approximately the same over the volume of the planet, 
and the pressure $P$ can be considered as a function of 
density $\rho$ only,
$P=P(\rho)$.  The perturbations of the planet may be
considered as adiabatic, and therefore 
the same functional dependence 
$P=P(\rho)$ holds during  perturbation as well. 
This condition 
is often referred to as the barotropic condition.  
In the  barotropic approximation the linearised 
Euler equations take the form (see PI)
\begin{equation}
\frac{\partial {\bf  v}} {\partial t }+2{\bmth {\Omega}}\times 
{\bf v} =-\nabla W + \frac{{\bf f}_{\nu}}{\rho}, \label{eq p3}  
\end{equation}
where
\begin{equation}
W=c_{s}^{2}\rho^{'}/\rho+\Psi_{ext},  \label{eq p4}
\end{equation}
${\bf v}= (v_r, v_{\theta}, v_{\phi})$ is the Eulerian velocity perturbation,
$\rho^{'} $ is the Eulerian
density perturbation, $c_{s}= \sqrt{dP/d\rho}$ is the 
adiabatic sound speed,  ${\bf f}_{\nu} $ is the viscous
or diffusive force pr unit volume and  as indicated above $\Psi_{ext}$ is 
the external forcing  tidal potential.  The velocity perturbation ${\bf v}$
is related to 
${\bmth{ \xi}}= (\xi_r, \xi_{\theta},\xi_{\phi}), $  the Lagrangian displacement vector, through
\begin{equation}
\frac{\partial {\bmth{\xi}}} {\partial t } = {\bf v}\label{xidot}.
\end{equation}
\noindent The linearised continuity equation gives    
\begin{equation}
P' =\rho^{'}c^2_s=-c^2_s\nabla \cdot (\rho {\bmth {\xi}}),  \label{eq p6}
\end{equation}
where $P'$ is the Eulerian  pressure perturbation.
Note that the centrifugal term is absent in equation  $(\ref{eq p3})$
being formally incorporated into the potential governing
the static equilibrium of the unperturbed star
and  there is no unperturbed motion in the rotating frame.

As in our previous work (eg. Papaloizou $\&$ Pringle 1981, PI, IP)
we shall neglect centrifugal distortion of the basic equilibrium  which will
enable us to adopt a spherically symmetric unperturbed  model
and density distribution. We additionally simplify matters
by adopting the Cowling approximation which neglects perturbations
to the planet gravitational potential, being equivalent to regarding
the planet as moving in a fixed specified spherical potential.
Although we shall adopt a standard polytrope with $n=1$ as the basic unperturbed model
which is not very centrally condensed, the main focus is on the low frequency
inertial modes for which self-gravity is not expected to play
a very important role. This is especially the case for small scale
modes and waves of the type that arise when a solid core  is considered.
The Cowling approximation should also be valid when estimating the effects
of compressibility on these kinds of disturbances.

\begin{figure}
\epsfig{file=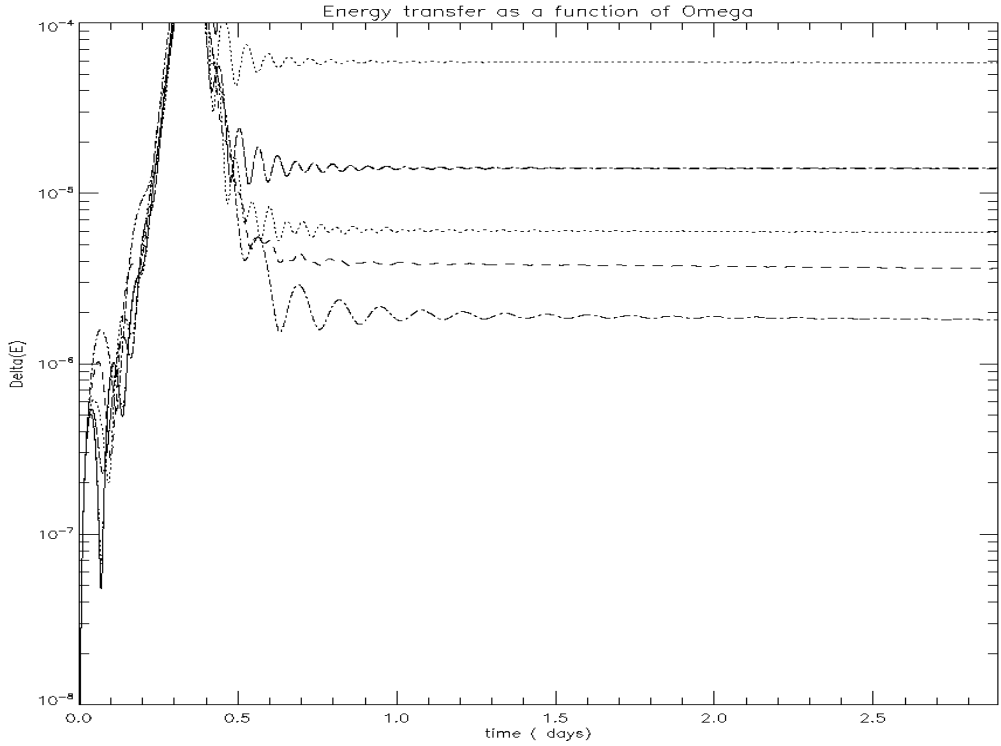,height=10cm, width=14cm}
 \caption{ The energy { measured in the rotating frame} transferred to the planet as a
  function of time  due to the  $m=2$  tide 
is plotted for different angular velocities
for an encounter with $\eta =4.$
The  curves from the uppermost to lowermost at large times
are for the non rotating
case,   $\Omega/\Omega_* = 0.16,$
$\Omega /\Omega_* = 0.36$ 
( the curve obtained with twice the numerical resolution 400*400 is superposed
and can barely be distinguished),
$ \Omega /\Omega_* = 0.51$,
$ \Omega/\Omega_*  = 0.8$
and $ \Omega/\Omega_*  = 1.12$ respectively
(the latter case was considered for the purposes of illustrating trends).
In this and  other similar  Figures  below the time unit is days 
and the energy is expressed in units of the natural
energy scale of the planet $E_*$. 
\label{fig1}}
\end{figure}

\begin{figure}
\epsfig{file=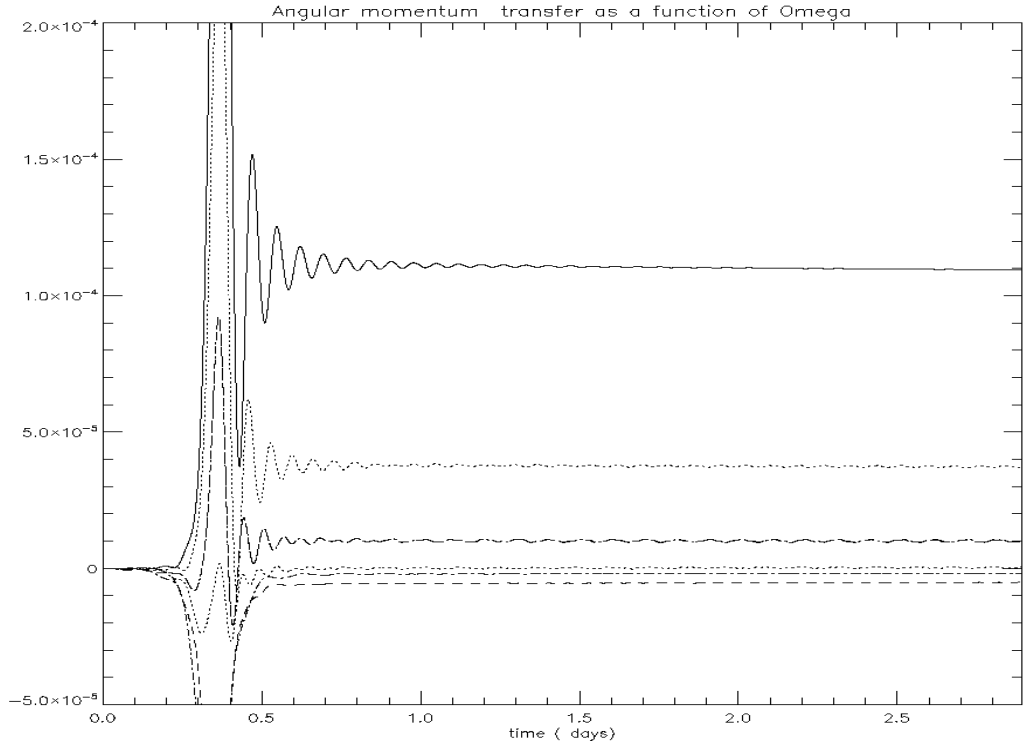,height=10cm, width=14cm}\caption{ 
The angular momentum transferred to the planet  induced by the $m=2$
tides is plotted as a function
of time for different angular velocities
for an encounter with $\eta =4.$
The  curves from the lowermost to uppermost at large time
are for $\Omega/\Omega_*  = 1.12,$
$\Omega/\Omega_*  = 0.8$, 
$ \Omega /\Omega_* = 0.51$,
$ \Omega /\Omega_* = 0.36$ (curve for $400\times400$
resolution superposed),
$ \Omega/\Omega_*  =  0.16$ and
 $\Omega =0$ respectively.
In this and other similar  Figures  below the time unit is days
and the angular momentum is expressed in units of the 
angular momentum  scale of the planet $L_*$.
\label{fig2}}
\end{figure}

\begin{figure}
\epsffile{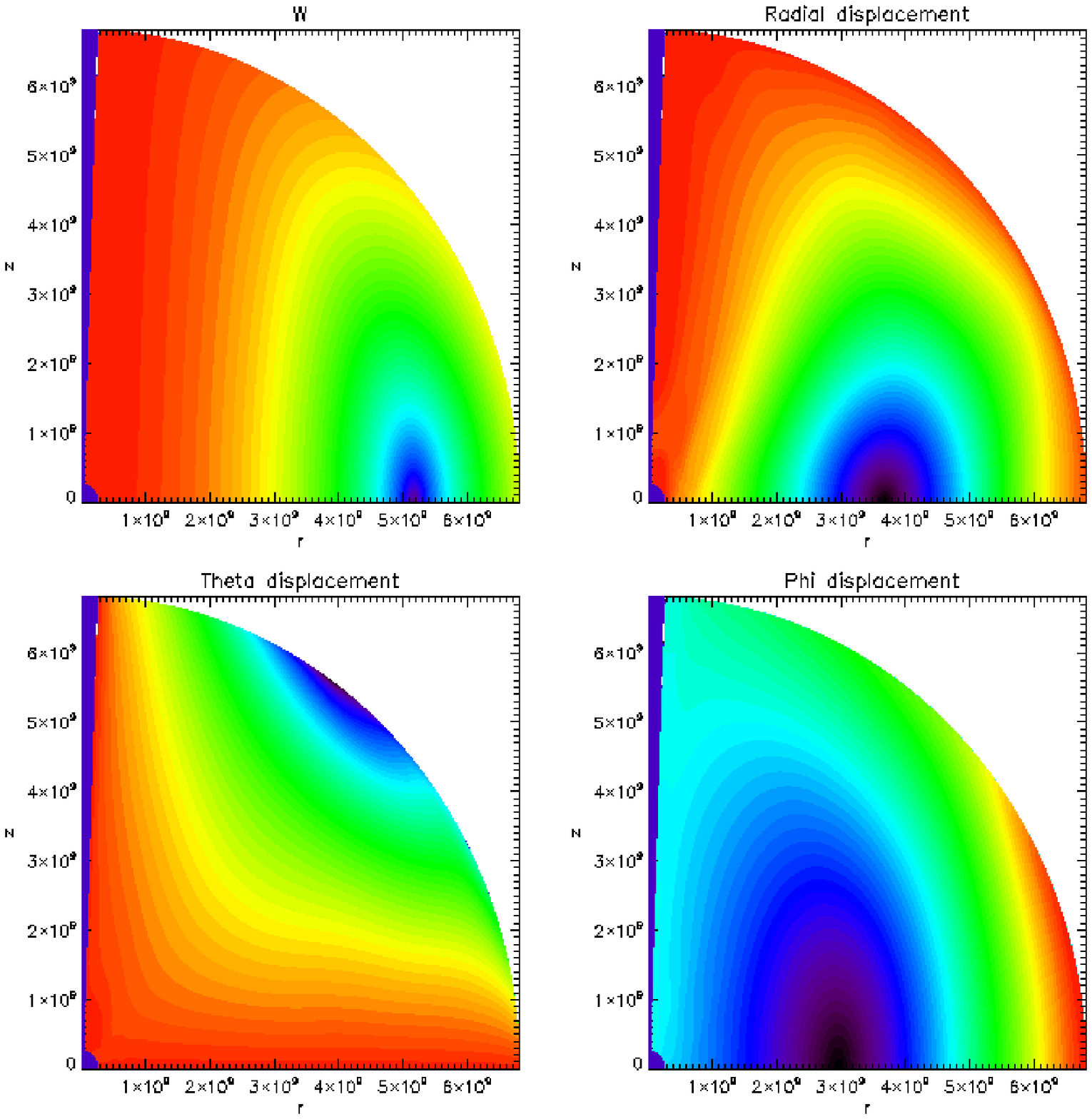}
 \caption{ Contour  plots  for the real parts of the $m=2$
 Fourier components of  $W$ (upper left panel),
 $\xi_r$ (upper right panel), $\xi_{\theta}$(lower right panel)
 and $\xi_{\phi}$ (lower right panel)
 after $t =  2.523 days$
 for the model with $\eta=4$ and $\Omega/\Omega_* =0.36 .$
 In this and other similar Figures, lengths are expressed in $cgs$
 units. For reference $R_*=7\times 10^9.$
 \label{fig3}}
\end{figure}

Provided that the expressions for the density,  sound speed 
are specified for some unperturbed model of
the planet as well as the form of the 
diffusive forces 
(see below) the set of equations  $(\ref{eq p3}-\ref{eq p6})$ is complete.
Details of its numerical solution are given in appendix A.

\subsection{Addition of diffusive effects}\label{sec2.2}
 
Although problems of the type considered
here are linear, because the spectrum is 
in a general sense singular,
arbitrarily small scales can develop over time.
It is important to note that such behaviour has a physical
basis and does not arise from pure numerical artifacts.

For example the boundaries of the simulations
we perform consist of the polar axis, the equatorial plane and
the surface of the planet. Because of the coordinate
singularity at $\theta=0,$ the polar axis is replaced by the cone
$\theta=\theta_{n_{th}} ,$ see below,   the latter angle  being 
small. The angle between this cone and the equatorial plane
is accordingly slightly less than a right angle.
As shown by Ralston (1973), inertial waves with an appropriate frequency
can undergo a sequence of alternating
reflections at these boundaries, resulting in  wave action
accumulating in the corner, where the boundaries  intersect.
Hence a reflectosingularity is produced implying that such frequencies
are in the continuous rather than point spectrum.
Note that even small perturbations to boundaries can introduce 
effects of this type.

In order to avoid numerical problems arising from such phenomena
we incorporate a diffusivity in the simulations which is adequate
to effectively  eliminate such artifacts in models with no specified central cores.

We added simple diffusive forces per unit
volume in spherical coordinates of the form 
\begin{equation}
{\bf f}_{\nu} = \frac{\partial}{\partial r}\left(\rho\nu\frac{\partial {\bf v}}{\partial r}\right)
+\frac{1}{r^{2}}
\frac{\partial }{\partial \mu}\left(\rho \nu\sin\theta \frac{\partial {\bf v}}{\partial \mu}\right),
\end{equation}
where the diffusion coefficient or effective kinematic viscosity
$\nu = \nu_0 \sqrt{GM_*R_*}$ was taken to be a constant. For the simulations
reported here with coreless models and a grid resolution of $200\times 200$
(for more information about the numerical grid see appendix A),
we adopted
$\nu_0=9.36\times10^{-6}$, and for these models with a grid
 resolution of $400\times400$
we adopted $\nu_0=4.68\times 10^{-6}.$
For models with core radius $0.25R_*$ we adopted $\nu_0=4.57\times10^{-5}$
at a grid resolution of $200\times 200$ and $\nu_0=2.28\times10^{-5}$
at a grid resolution of $400\times 400.$ 
Finally, for models with core radius $0.5$ we adopted $\nu_0=1.7\times10^{-5}$
at a grid resolution of $400\times 400$ and $\nu_0=8.5\times10^{-6}$
at a grid resolution of $800\times 800.$  With these choices
the diffusivity decreases as the resolution increases.
We remark that energy and angular momentum transfers measured
by evaluating the canonical energy and angular momentum
(see appendix A)  just after the encounter
are robust to changes in numerical resolution and diffusivity as they should be
(see below).

\begin{figure}
\epsfig{file=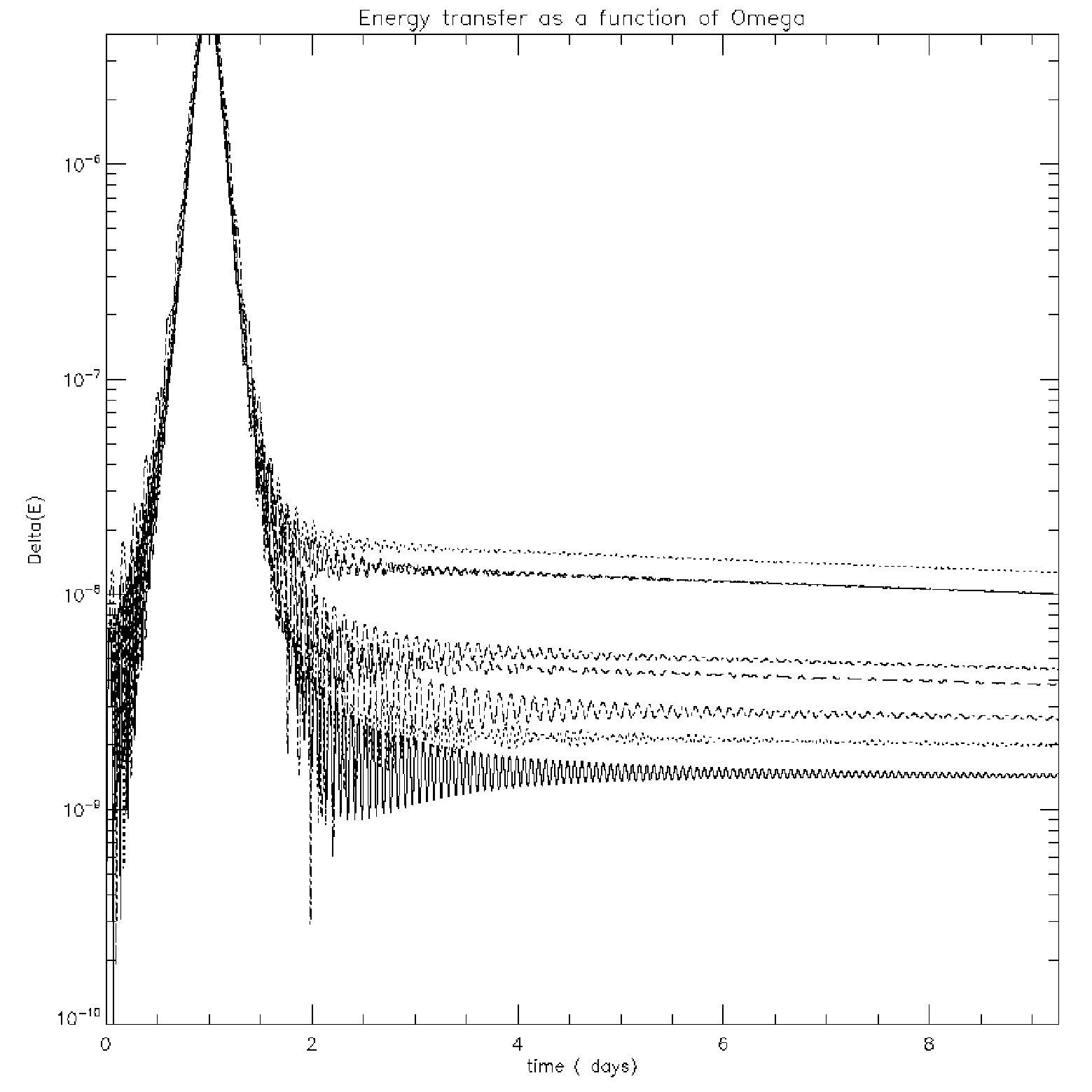,height=10cm, width=14cm} \caption{
The energy transferred to the planet as a function of time due  $m=2$  tide
is plotted for different angular velocities
for an encounter with $\eta =8\sqrt{2}.$
The  curves from the uppermost to lowermost at large time
are for $\Omega /\Omega_*= 0.36,$
$\Omega/\Omega_* = 0.27$
(the curve obtained for $\Omega/\Omega_* =0.45$ overlaps),
$ \Omega/\Omega_* = 0.6$,
$ \Omega/\Omega_* = 0.17,$
$ \Omega /\Omega_*= 0.8,$
$ \Omega /\Omega_*= 0.1$ and
$\Omega/\Omega_* =0$, respectively.
\label{fig4}}
\end{figure}

\begin{figure}
\epsfig{file=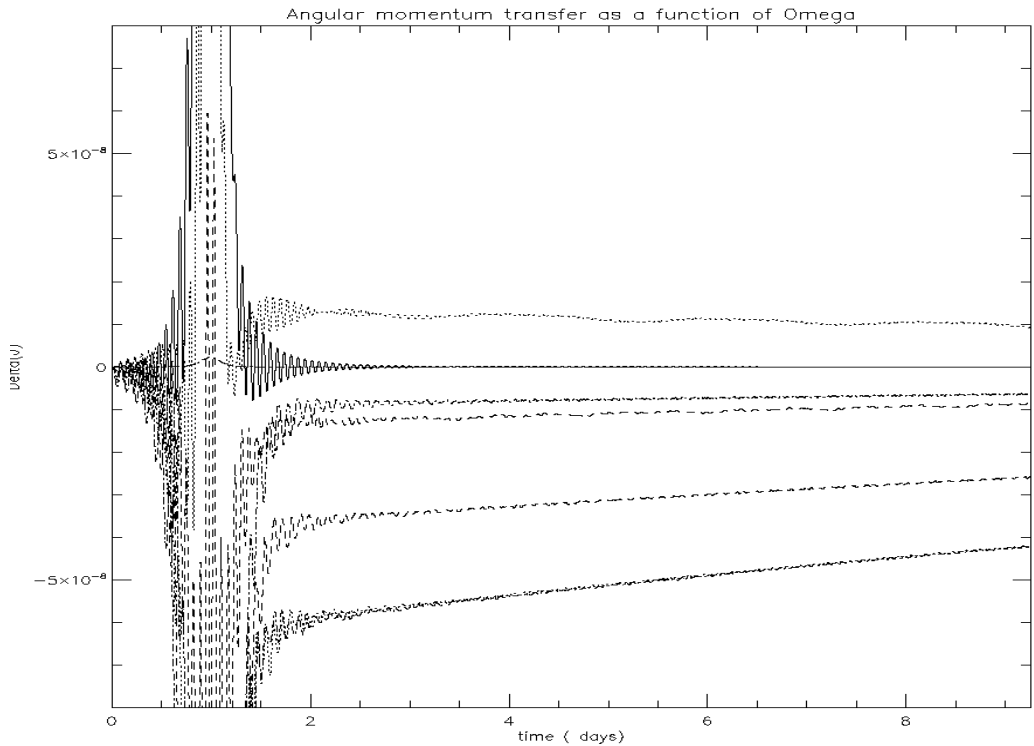,height=10cm, width=14cm} \caption{
The angular momentum transferred to the planet is plotted
as a function of time for different angular velocities
for an encounter with $\eta =8\sqrt{2}.$
The  curves from the uppermost to lowermost at large time
are for $\Omega /\Omega_*= 0.1,$
$\Omega =0,$
$ \Omega/\Omega_* = 0.8,$
$ \Omega /\Omega_*= 0.6,$
$ \Omega/\Omega_* = 0.17,$
$ \Omega/\Omega_* = 0.45$ and
 $\Omega/\Omega_* = 0.27$
( the curve obtained for $\Omega/\Omega_* = 0.36$ overlaps),
 respectively.
 \label{fig5}}
\end{figure}

\begin{figure}
\epsfig{file=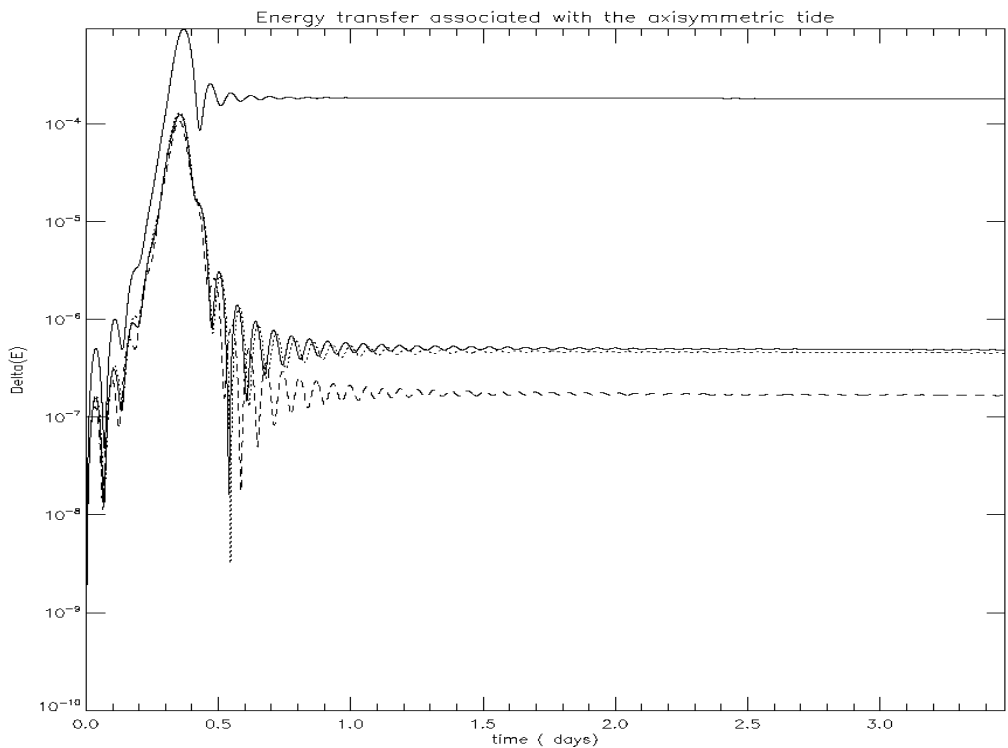,height=10cm, width=14cm}
\caption{
The energy transferred to the planet as a function of time due to the
axisymmetric tide $(m=0)$ is plotted for different angular velocities
for an encounter with $\eta =4.$ The lowermost curve at large times
corresponds to $\Omega/\Omega_* = 0.8,$ the
curve immediately above this to $\Omega/\Omega_* =0.36$ and
curve immediately above that , which is almost coincident,  is for the non rotating case.
For purposes of comparison the uppermost  curve is for
the non rotating  case  with the $m=2$  component of the
tidal potential.
 \label{fig6}}
\end{figure}

\begin{figure}
\epsfig{file=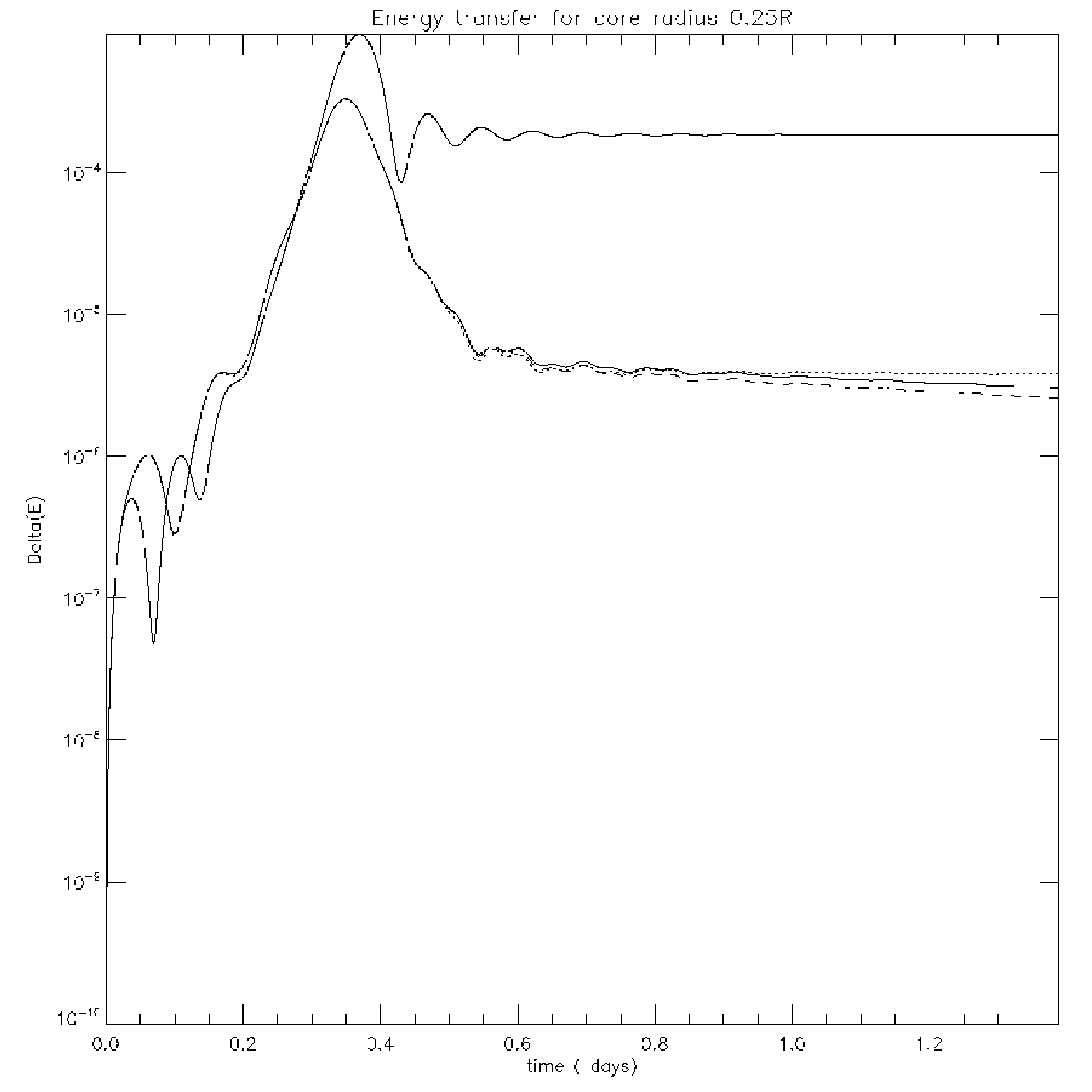,height=10cm, width=14cm}
\caption{The energy transferred to the planet as a function of time due to the  $m=2$  tide
is plotted for  a model with a solid core of radius $0.25R_*.$
The  encounter is for $\eta =4$ and
 $\Omega/\Omega_* = 0.8.$
 The  curves from the lowermost to uppermost at large time
are for a resolution of $200\times 200,$
a resolution of $400*400,$ for comparison a model with no core
and a model with no core and no rotation.
\label{fig7}}
\end{figure}

\begin{figure}
\epsfig{file=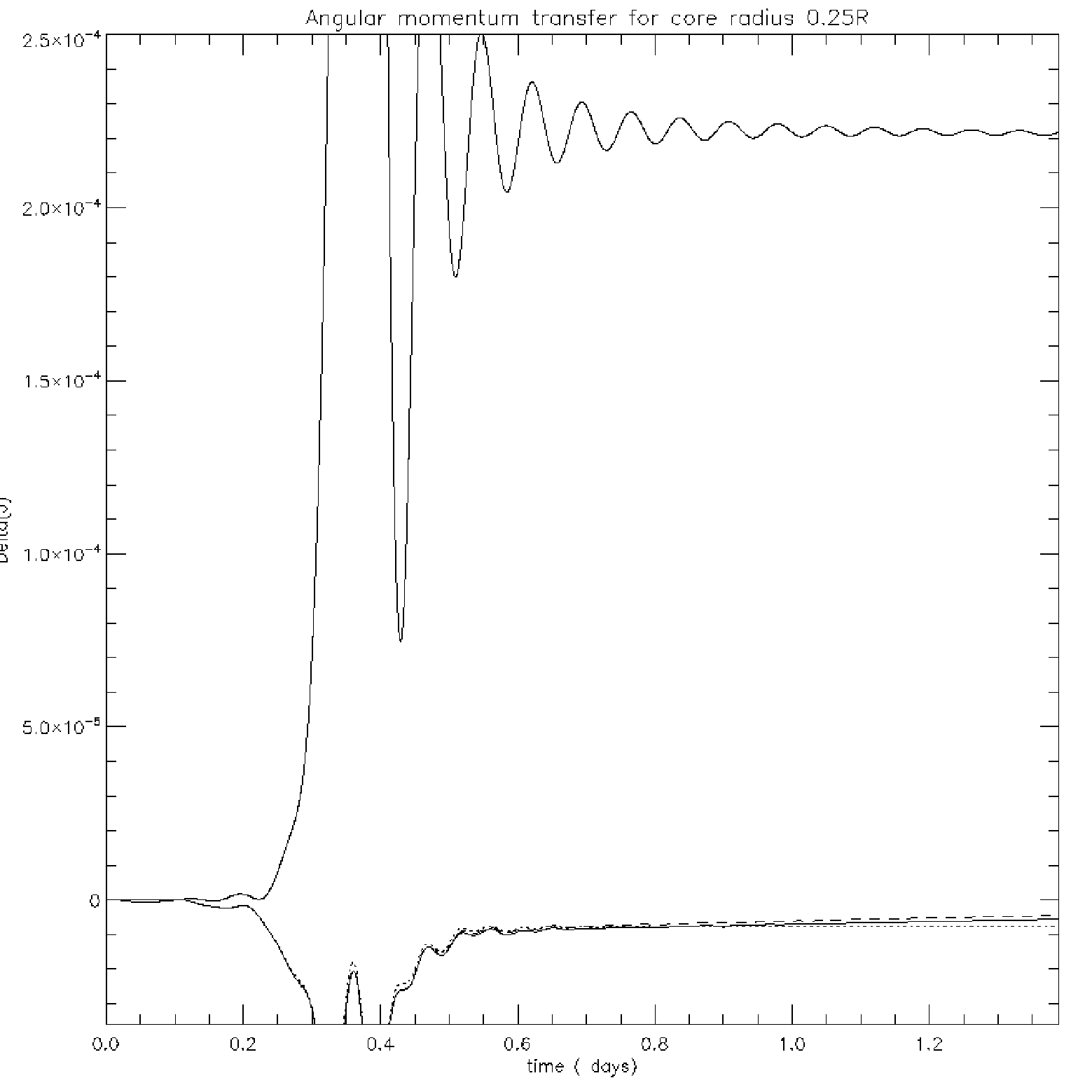,height=10cm, width=14cm}
\caption{
The angular momentum transferred to the planet as a function of time due to the  $m=2$  tide
is plotted for  a model with a solid core of radius $0.25R_*.$
The  encounter is for $\eta =4$ and
$\Omega /\Omega_*= 0.8.$
The  curves from the lowermost to uppermost at large time
are for a resolution of $800\times 800,$
a resolution of $400*400,$ for comparison a model with no core
and a model with no core and no rotation.
\label{fig8}}
\end{figure}

\section {A general description of numerical results}

\subsection{The energy and angular momentum transfer for a close encounter}

We here give results for the energy and angular momentum transfer as a result of 
a tidal encounter at fixed pericentre distance for different angular velocities.
We have considered a range of $\Omega/\Omega_*$, where $\Omega_{*}$ is defined in
equation (\ref{eqn p2}), up to $1.12$ even though
we have neglected centrifugal terms in the model.  This is done in order to 
more fully explore
and gain understanding of  the functional dependence 
 of the response on its  defining  parameters.  The introduction
of centrifugal terms would not change the essential response problem
although deviations of the form of the effective surface  boundary may play some role (see eg. discussion in section \ref{sec2.2} above).

A first set of runs was carried out for $\eta=4$ taking into account the $m=2$ component of the tidal forcing.
The contribution from this has found to dominate that of the $m=0$ component (see below).
In all of these and other similar cases  with no interior solid core, after the encounter
 the canonical energy and angular momentum attain
either expected constant values or values that slowly decay on account
of the imposed  numerical diffusion. We have considered resolutions of $200\times 200, $
$400\times 400$ and for models with solid cores $800\times 800.$
As illustrated below the energy and angular momentum transferred between central
object and planet is
 measured just after the encounter.  This measures the physical
exchange  and is numerically converged.

The energy  transferred to the planet as a function of time  due  to the  $m=2$  tide 
is plotted for different angular velocities
for an encounter with $\eta =4$
in Figure \ref{fig1}.  Results  are given for  the  non rotating
case,   $\Omega/\Omega_*= 0.16,  0.36,  0.51, 0.8$
and $ \Omega/\Omega_* = 1.12$, respectively.
The angular momentum transferred to the planet as a function
of time for the cases illustrated in Figure \ref{fig1}
are plotted in Figure \ref{fig2}.
The energy transferred is seen to decrease monotonically
with angular velocity by about a factor of thirty
as $\Omega$ varies between zero and the maximum value considered.
The energy transferred is expected to decrease as the star is spun up from zero
(Ivanov \& Papaloizou 2004) but may increase again at large values 
of $\eta$ (see  PI, IP and below).

The angular momentum transferred  to the planet  has a maximum  absolute value
for the non rotating model and decreases with increasing angular velocity, becoming  zero
when $\Omega/\Omega_* = 0.51.$  For larger values of $\Omega$  angular momentum is transferred
from the planet to the star.  Such a sign change is indeed 
expected when  $\Omega$ is comparable
to the angular velocity at pericentre, see PI and IP.
We remark that the simulations for $\Omega/\Omega_* = 0.36$ were done 
at resolutions of $200\times 200$ and $400\times400$ and are
 illustrated in Figures \ref{fig1}
and \ref{fig2}.  The results  of these  are indistinguishable.

The form of the planet response in these cases is found to be
global. To illustrate this, contour  plots  for the real parts of the $m=2$
 Fourier components of  $W,$
 $\xi_r,$  $\xi_{\theta}$
 and $\xi_{\phi}$ (lower right panel)
 after  $t =  2.523 days,$ well after the encounter is complete,
 for the simulation with  $\Omega/\Omega_*= 0.36$ are shown in Figure \ref{fig3}
 These have global form that is  typically found at any time.  As expected, there is no evidence
 of evolution to small scale structures indicative of { the emission of shear layers
 at critical latitudes and } wave attractors 
 (eg. Ogilvie \& Lin 2006, { Rieutord \& Valdettaro 2010})
 in these cases. The global form is indicated both by the  accurate conservation
 of canonical energy and the accurate conservation  angular momentum even though some numerical diffusion
 was present and also that the energy and angular momentum transferred is
 in good agreement with values obtained using a basis function approach
developed in our previous papers PI and IP 
with a relatively small number $15\times 15$ of trial functions, see the next section.

\subsection{The energy and angular momentum exchanged for a more distant encounter}

We  now consider simulations of the above form but with a larger value of  
 $\eta =8\sqrt{2}.$
In addition to the non rotating case,  the  angular velocities considered  were 
$ \Omega/\Omega_* = 0.36, 0.27,$
$0.45,0.6,$
$0.17, 0.8$ and 
 $0.1.$
The energy transferred to the planet as a function of time is illustrated
in Figure \ref{fig4},
and the angular momentum transferred to the planet is illustrated in Figure \ref{fig5} .
In this case the energy transferred initially increases with angular velocity
but attains a maximum value before decreasing again.
The reason for this behaviour is that the $f$ mode is ineffective at this
value of $\eta$ leading to a weak interaction for the non rotating model.
However, once the star is spun up, the interaction can strengthen
through the inertial  mode response.
In this case the maximum transfer  occurs for $\Omega /\Omega_*\sim 0.36.$ 
Similarly the angular momentum transfer is very small  for the non rotating model
It increases to a maximum for $\Omega/\Omega_*=0.1$ and then passes
through zero at $\Omega/\Omega_* \sim 0.13.$ For larger angular velocities
angular momentum is transferred from the planet to the star, this transfer being
largest in magnitude when $\Omega/\Omega_*  \sim 0.32$ beyond which it decreases
in magnitude. The sign change again occurs at a value of the angular
velocity comparable to that at pericentre passage and the planet response remains global.

\begin{figure}
\epsffile{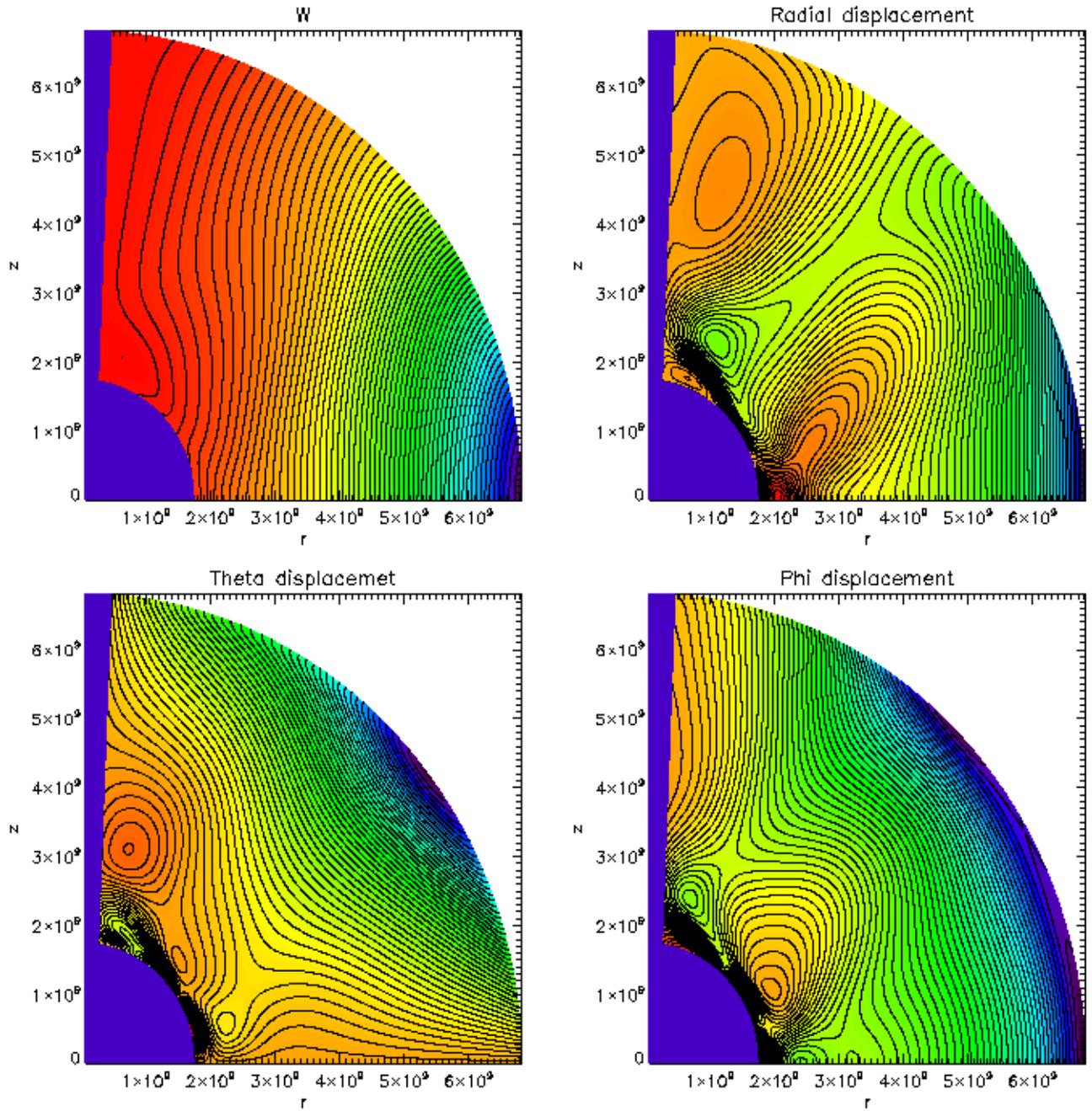}
\caption{ As in Figure \ref{fig3}
but for a model with core radius of $0.25R_*$
and $\Omega/\Omega_* = 0.8$
 after $t =   0.408 days.$
 \label{fig9}}
\end{figure}

\begin{figure}
\epsffile{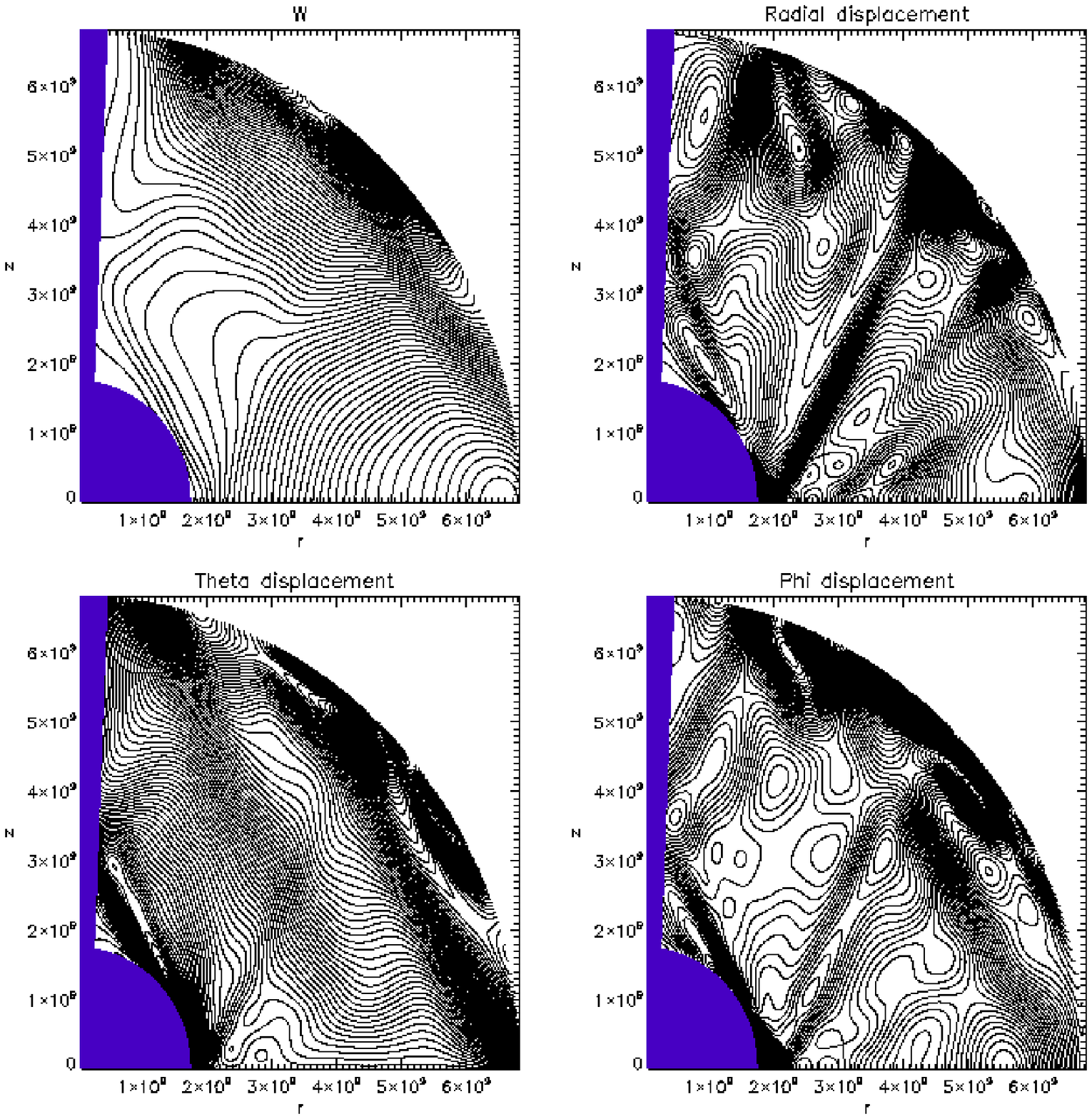}
 \caption{ As in Figure \ref{fig9}  but  at $t =  0.776 days$
 and with contour line plots
 in order to illustrate the inertial wave propagation pattern
 \label{fig10}}
\end{figure}

\subsection{The axisymmetric case $m=0$ for $\eta = 4$}

In order to evaluate the effectiveness of the $m=0$ tide we 
carried out simulations using the $m=0$ component of the tidal potential
for $\eta=4.$
The energy transferred to the planet as a function of time is plotted for 
$\Omega/\Omega_* = 0.8,  0.36$ and the non rotating case in Figure \ref{fig6}.
The non rotating case for $m=2$ is also shown. The energy transfer for $m=0$
is always found to be an order of magnitude or more smaller than the
corresponding  $m=2$ contribution. Therefore this can be neglected.

\subsection {Planet models with a solid core}
\subsubsection{A planet with core radius equal to $0.25R_*$}

\begin{figure}
\epsfig{file=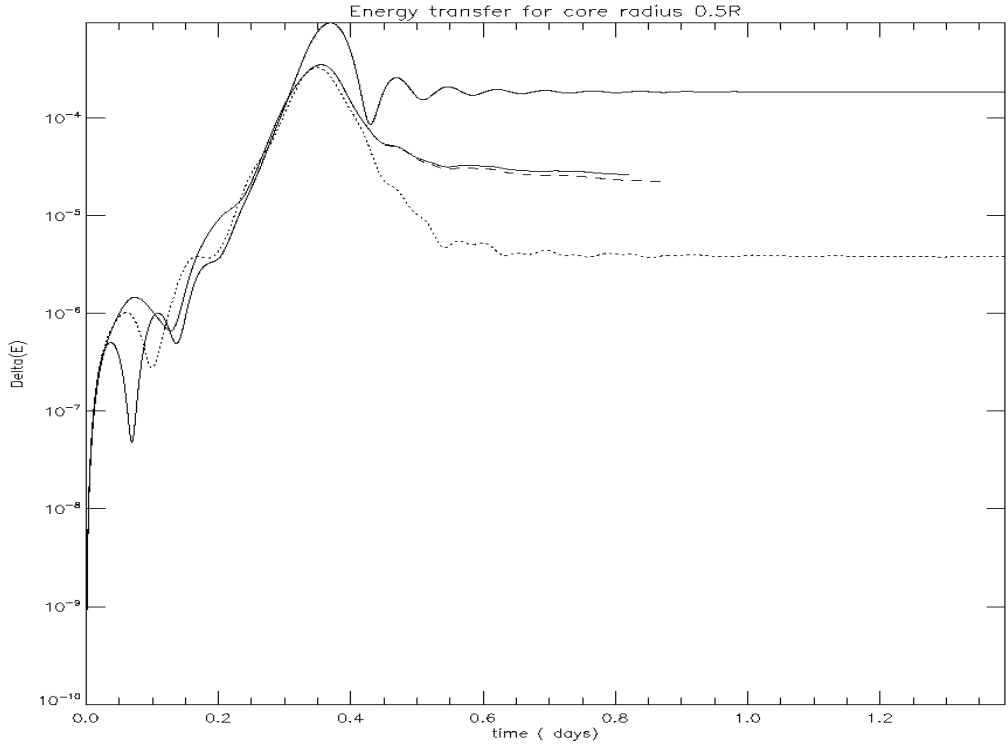,height=10cm, width=14cm}
\caption{
The energy transferred to the planet as a function of time due to the $m=2$  tide
is plotted for  a model with a solid core of radius $0.5R_*$.
The  encounter is for $\eta =4$ and
 $\Omega/\Omega_* = 0.8.$
 The  curves from the lowermost to uppermost at large time
are for a comparison model with no core,  for a resolution of $400\times 400,$
for a resolution of $800*800,$ for further  comparison a model
 with no core and no rotation.
\label{fig11}}
\end{figure}

We now go on to consider planet models with a solid core.
In these cases the long term behaviour differs from 
the coreless models  owing to
the development of wave patterns related to { critical latitude phenomena and } wave attractors
(eg. Ogilvie \& Lin 2006, Ogilvie 2009 and references therein) that correspond
to the development of increasing concentrations
of wave energy along closed characteristic paths.
 However, it is important to note
that our calculations are not for fixed frequency forcing,
which is the situation considered for the discussion of  critical latitude phenomena and  wave attractors.
Here we are concerned with a combination of different frequencies
that should lead to a broader pattern. However, we still might expect that
singular behaviour develops asymptotically for large time, in the non
diffusive  case. In the diffusive case this would be limited but
energy dissipation would occur at an enhanced rate
as compared to the non singular coreless models. This feature is observed
in our simulations.

\begin{figure}
\epsfig{file=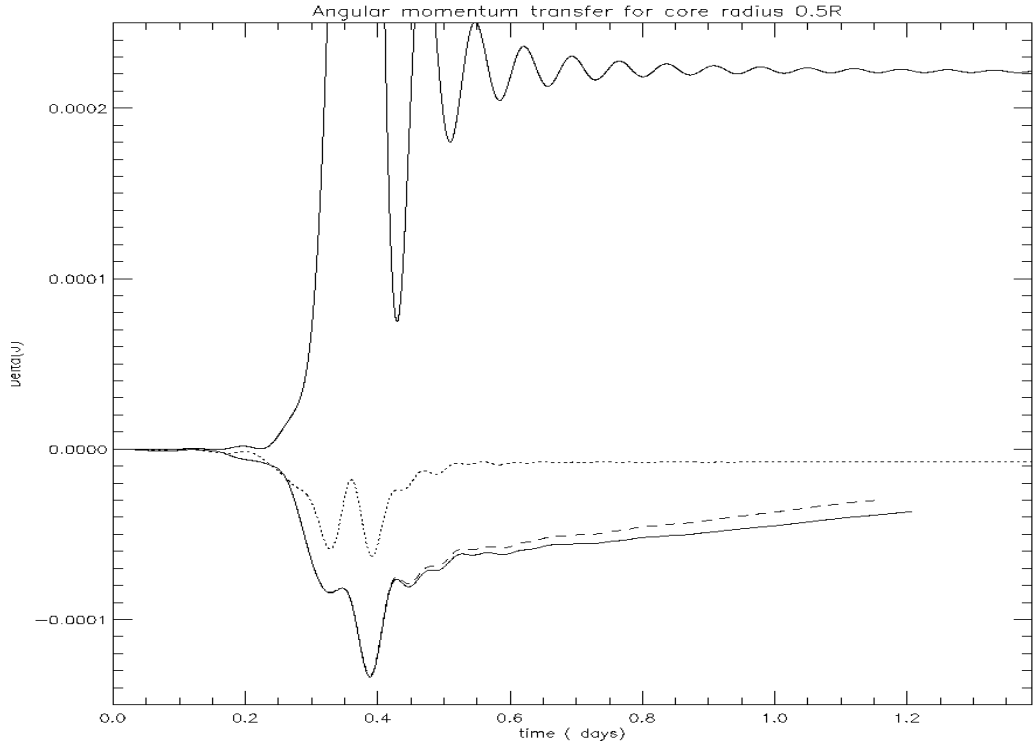,height=10cm, width=14cm}
 \caption{
 The angular momentum  transferred to the planet as a function of time 
due to the $m=2$  tide 
is plotted for  a model with a solid core of radius $0.5R_*.$
The  encounter is for $\eta =4$ and
 $\Omega/\Omega_* = 0.8.$
 The  curves from the lowermost to uppermost at large time
are   for a resolution of $800\times 800,$
for a resolution of $400*400,$ for the corresponding model with no core
  and  for further  comparison a model
 with no core and no rotation. \label{fig12}}
\end{figure}

\begin{figure}
\epsffile{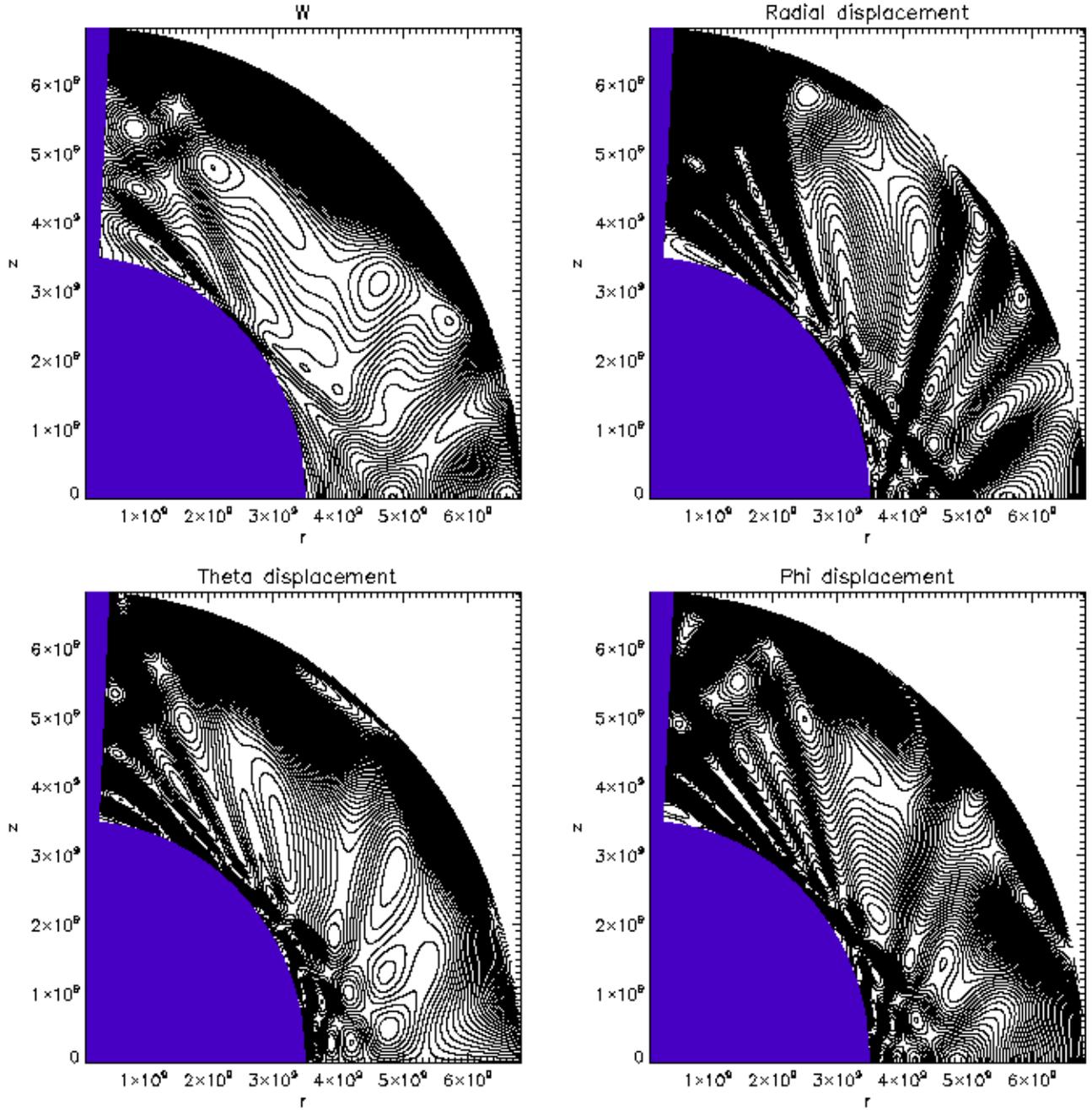}
\caption{As in Figure \ref{fig10} but
for the model with a  core radius of $0.5R_*$ at $t = 0.589 days.$
 \label{fig13}}
\end{figure}

\begin{figure}
\epsffile{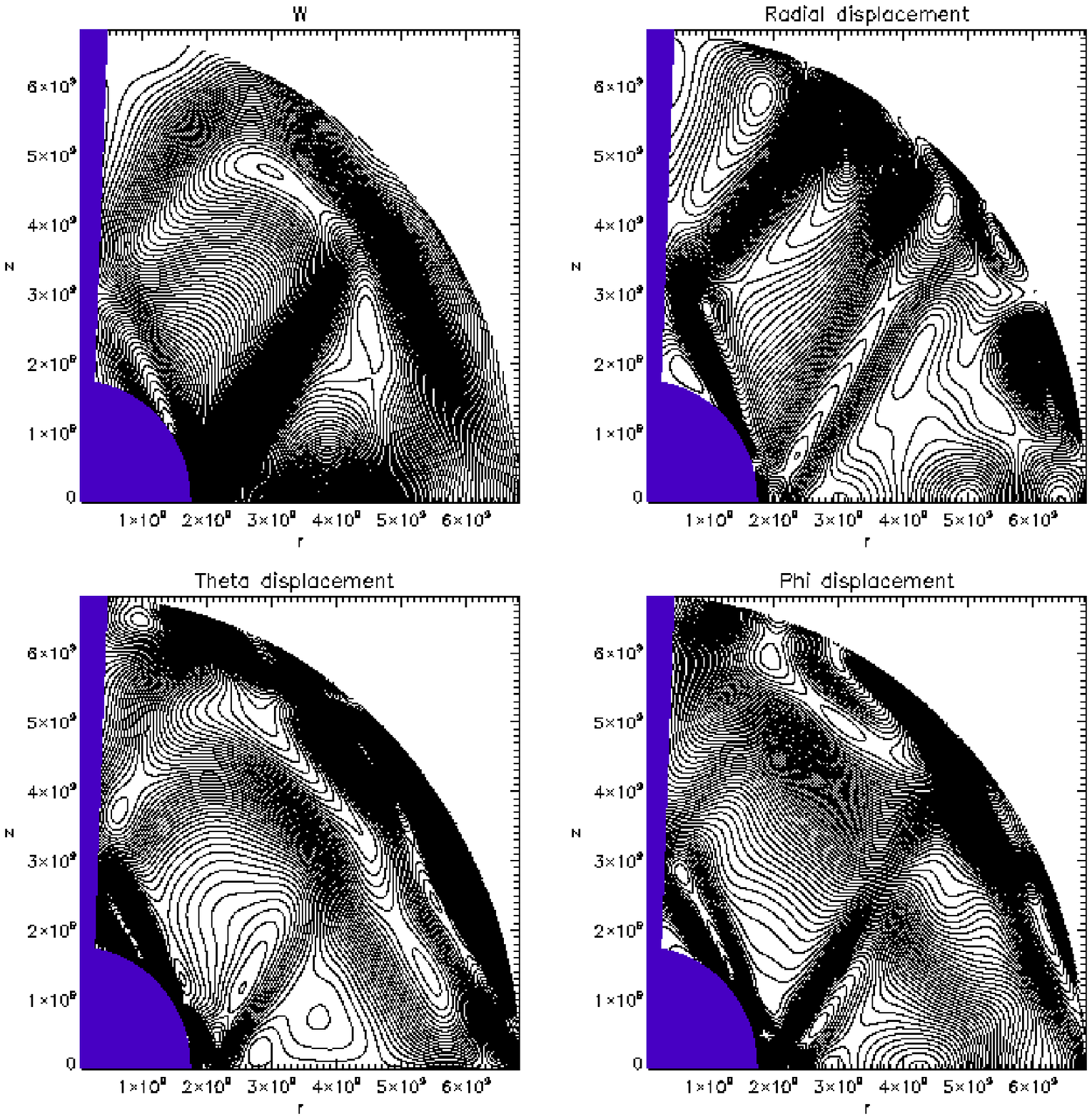}
 \caption{ As in Figure \ref{fig9}  but for a model with core radius
$0.25R_*,$
 $\Omega/\Omega_* = 0.36 $ and $\eta=8\sqrt{2}$  at $t =  2.326 days$,
 which is after the encounter. Contour line plots
  illustrate the inertial wave propagation pattern
 \label{fig10a}}
\end{figure}

The energy transferred to the planet as a function of time due to the  $m=2$  tide 
is shown in Figure \ref{fig7} for  a model with a solid core of radius $0.25R_*$
for $\eta =4$ and  
 $\Omega/\Omega_* = 0.8.$
Simulations have been performed 
with   resolutions  of $200\times 200$ and
 $400*400.$ The results agree closely with some small additional dissipation
in the lower resolution case, which is to be expected.
Interestingly, the energy transferred to the planet is
almost the same in the cases  with and without a core.
This transfer is less in magnitude than in the non rotating case, which
gives almost identical results  to those found for  the non rotating coreless case.

The corresponding angular momenta  transferred to the planet are
plotted as a function of time in Figure \ref{fig8}. As for the energy transferred, 
the angular momentum transferred in this model
 and the coreless model are very similar.

The global response pattern after $t=0.408 days$ is illustrated
in  Figure \ref{fig9}. 
 This  is shortly before the end of the  encounter.
Although there is evidence for some small scale structures developing
near the core boundary, the basic response appears global with  little
evidence of  { critical latitude phenomena or } wave attractors. However, this appears 
at later times. Response pattern contours are shown in
Figure \ref{fig10}  at  $t =  0.77{\rm  days},$
being  well after the end of the encounter.
Localised  inertial wave propagation and reflection  patterns 
 are clearly visible. As expected these present  a constant angle of propagation
to the vertical and graze the core boundary.

The fact that the development of these structures occurs at late times
is indicative of why the coreless and cored results agree in this case.
We note that this may be expected when the time scale for the encounter
is not long  compared to the rotation period. In such a case there would not be
adequate time for inertial waves to propagate, reflect  and  develop structures
indicating { shear layers or } attractors.  This might be expected to be especially the case
when the core is small and so presents a small target for propagating waves. 
Physically, as far as the inertial mode excitation is concerned,
the encounter would appear impulsive in which case the energy and
angular momentum transfers should not be affected
by the later development of { shear layers or  wave attractors.}  Therefore, as long as the
core is not too large we expect the same result as for  a coreless model.
These ideas are developed from a mathematical  point of view  
in  appendix B. 
Although the encounter time scale and rotation period are comparable
in this case, in accordance with the above discussion,
we do not see the development of wave propagation patterns
until the late stages of the encounter, and the cored and coreless model
energy and angular momentum  transfers
are very similar.


\subsubsection{A planet with core radius equal to $0.5R_*$}

We have also performed simulations   a model 
 with a solid core of radius $0.5R_*.$
Again the  encounter was for $\eta =4$ and  
 $\Omega/ \Omega_*= 0.8.$
The energy  and angular momentum transferred as a function of time 
are shown in Figures \ref{fig11} and \ref{fig12}, respectively.
Results are given for resolutions of $400\times 400$ and 
 $800\times 800.$ These agree closely but with again more dissipation
in the lower resolution case. In contrast to the case of the smaller core
the energy transferred is larger than in the coreless case
by a factor of $\sim 8-9.$ This is probably because the larger core
enabled a wave structure to develop faster.  
To illustrate the wave pattern,  contour plots of the response
are shown in Figure \ref{fig13}  at $t = 0.59 days$,
which is shortly after the encounter.
These appear to be of a similar form to, but somewhat more intricate than the model
with a smaller core. 
\begin{figure}
\vspace{2cm}
\epsfig{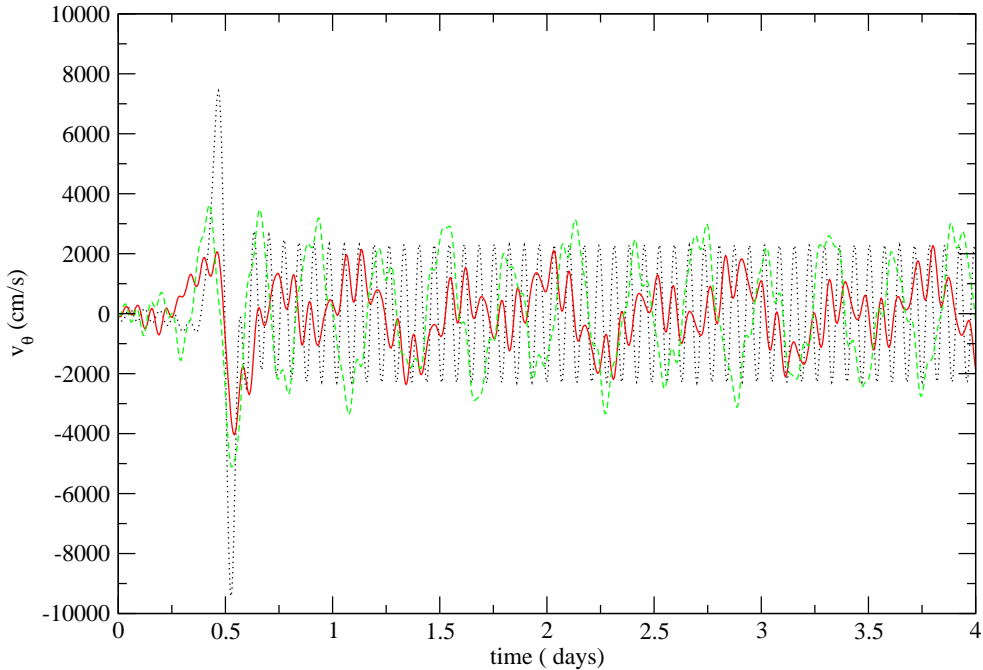}\vspace{1cm}\caption{
The real part  of the $\theta-$component of velocity in an inner point of
the planet as a function of time, $\eta=4\sqrt{2}$ and $m=2$
 for  these tidal encounters.  Simulations for
$\Omega/\Omega_*=0, 0.24$ and $0.36$   are plotted with  dotted,
solid and dashed curves, respectively.
\label{fn1}}
\end{figure}
\begin{figure}
\vspace{2cm}
\epsfig{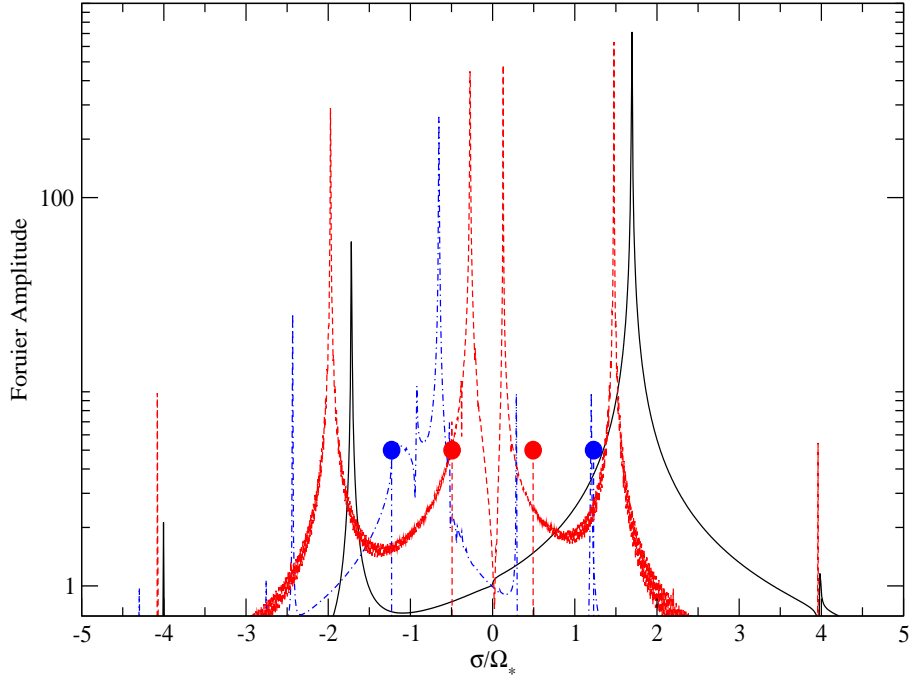}\vspace{1cm}\caption{ 
Fourier spectra obtained by the Fourier transform of the velocity data as functions of the
ratio $\sigma/\Omega_{*}$  for  $\eta=4\sqrt{2}.$
 The case $m=2$ is shown. 
The solid (black) curve corresponds to the non-rotating case $\Omega=0$ while
the dashed (red)  and dash-dotted (blue)  curves are for $\Omega/\Omega_*=0.24$ and
$0.6$  respectively. The vertical lines delimit the range of
$\sigma $, where  inertial modes are  theoretically possible, with 
 line styles identical to those used for
corresponding 
values of $\Omega.$
\label{fn2}}
\end{figure} 
\begin{figure}
\vspace{2cm}
\epsfig{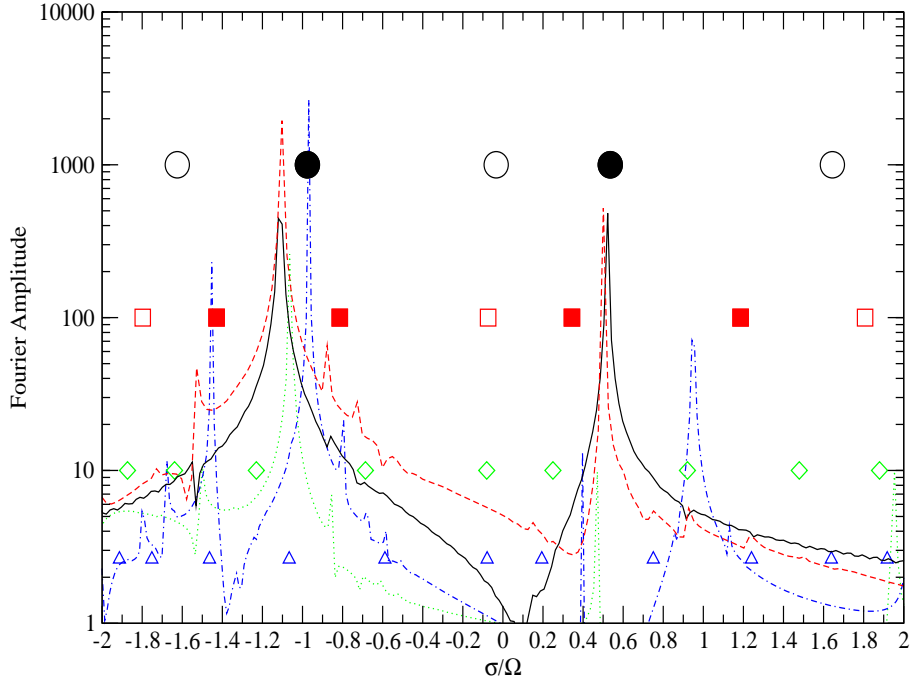}\vspace{1cm}\caption{ 
The Fourier spectra as functions of the ratio $\sigma /\Omega $,
with only the range corresponding
to the inertial branch of pulsations $|\sigma /\Omega| \le 2$ explicitly shown, for $m=2$ and 
$\eta=4\sqrt{2}$. The solid (black), dashed (red) , dotted (green) and dot-dashed (blue) curves are for  
$\Omega/\Omega_*=0.24, 0.36, 0.6 $ and $1.0.$ 
 Note that  { the latter case}
 is  unrealistic but included for completeness.  The symbols show positions
of the WKBJ modes obtained in IPN, with different style corresponding to different WKBJ mode order, 
$l$. The circles, squares, diamonds and triangles are for $l=1,2,3$ and $4$, respectively.
\label{fn3}}
\end{figure} 
\subsubsection{Cored models for slower rotation and larger $\eta$}
In order to indicate that  the results described above
are not radically altered when $\Omega$ is reduced and $\eta$ is increased,
we also simulated cored models with $\eta=8\sqrt{2}$ and 
 $\Omega/\Omega_* = 0.36.$ The models with core radii $0.25R_* $
and $0.5R_*$ behaved very similarly, in relation to their coreless
counterparts,  to those described above for larger $\Omega$ and smaller
$\eta.$ Figure \ref{fig10a} 
illustrates the wave propagation pattern that results from the encounter
 when the core radius
was $0.25R_*,$ which is indeed similar to that found for the cases
with larger rotation.

\section{A comparison of the numerical, spectral and WKBJ approaches }
\subsection{ Normal mode frequencies obtained from the simulations
and their comparison with the basis function and WKBJ approaches }

Here we investigate the spectrum of normal modes excited in the above simulations
determined from  fast Fourier transform of the time series.
We compare the derived inertial mode frequencies with the results obtained in IPN.  

As time series used for our Fourier analysis, for definiteness, 
we consider the time dependence of the real and imaginary parts of 
the ${\theta}-$component of velocity at an inner point of the planet,   with $r=4.8\times 10^9 cm$
and $\theta = 67$ degrees,
after a tidal encounter with $\eta=4\sqrt{2}$, for different values of the angular velocity $\Omega $. 
This  is shown in Fig. \ref{fn1} for $\Omega/\Omega_*= 0.24$ and $0.36$.  The  sharp change occurring
 around  $0.5$ day is due to the tidal encounter itself, the pulsation modes 
are generated during this period of time. After approximately $1$ day the influence of the perturbing 
tidal field becomes unimportant and the planet's perturbation becomes quasi-periodic. As follows from Fig. \ref{fn1}, 
in the case of the non-rotating planet,
  only one relatively short  period of pulsation can be clearly seen. 
This is the period of the fundamental prograde mode. When the planet rotates several periods can be seen with
one short  period corresponding to the fundamental mode together with several relatively long periods 
determined by presence of  inertial modes in the spectrum.

\begin{figure}
\vspace{2cm}
\epsfig{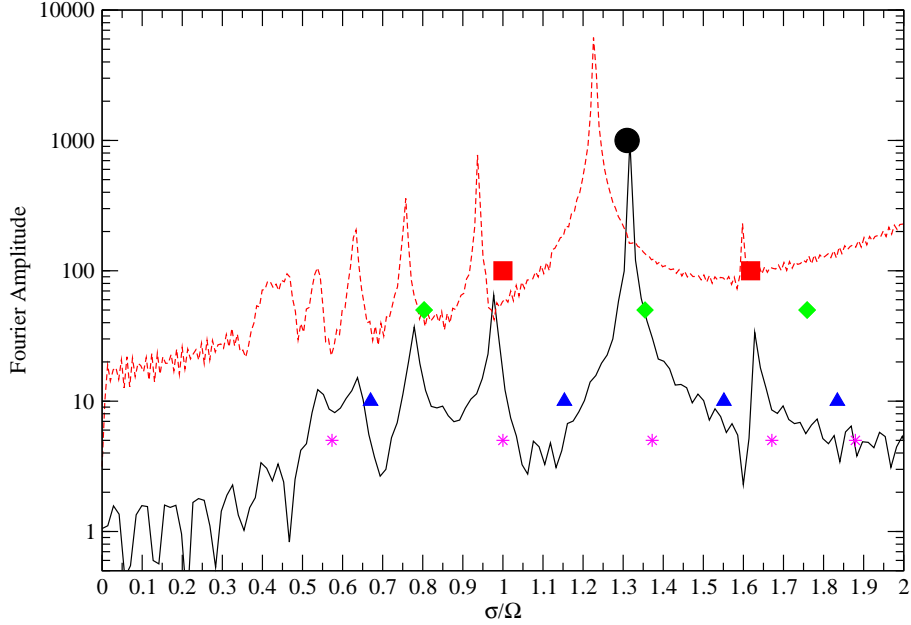}\vspace{1cm}\caption{ 
Same as Fig. \ref{fn3} but for the case $m=0$, $\eta=4$. 
Since in this case the spectra are symmetric about
the origin $\sigma=0$ only the positive values of $\sigma $ within the 
inertial range $\sigma/\Omega \le 2$
are shown. $\Omega/\Omega_*= 0.36$ for 
the solid (black) curve and $\Omega/\Omega_*=0.8$ for the dashed (red)  one.
Symbols show the positions of $WKBJ$ modes. 
Note that the WKBJ mode order $l$ runs starting from $l_{min}=0$ in
this case. The black circles, red squares, green  diamonds, blue triangles and  magenta stars 
are for $l=0,1,2,3$ and $4$, respectively.
\label{fn4}}
\end{figure}

To obtain the Fourier transforms  we removed the first part of the data
series to eliminate the contribution of direct  forcing by the
tidal field, typically of duration $5\times 10^4-10^5s$.   
The results are presented in Figs. \ref{fn2}-\ref{fn4}. In Fig. \ref{fn2} we compare the spectrum obtained for
the non-rotating planet with those obtained for the rotating cases, for  
$m=2$. The spectra are plotted as functions of 
the ratio $\sigma /\Omega_{*}$, where $\sigma $ is the frequency. The peaks in Figure \ref{fn2} correspond to positions 
of  normal modes. The black curve represents the results obtained for $\Omega=0$.  The positions of the two peaks are
symmetric about $\sigma=0,$ though the peak amplitudes are not. This is because 
the prograde  modes    with positive values of $\sigma $ 
are excited much more effectively than the retrograde modes,  for a non-rotating planet
\footnote{Let us recall that it is assumed in this paper that that the direction 
of planet's rotation and the orbital motion coincide and that  
planet's perturbations are proportional to $e^{-i(\sigma t - m\phi)}$,
where $\phi $ is the azimuthal angle in the rotating frame.
 It is easy to see that the modes with $\sigma > 0$ 
($\sigma < 0$) are prograde (retrograde) with respected to the orbital motion.}. The prominent peaks 
with $\sigma \approx \pm 1.6\Omega_{*}$ give the positions of the fundamental mode. Note that their 
eigenfrequencies are larger than those calculated for the same planet  model but without  the Cowling approximation. 
The peaks with larger values of $\sigma /\Omega_{*}$ give the  positions of p-modes 
(pressure waves). 
Note the absence of any peaks in between the two peaks corresponding to the fundamental mode. This is because 
there is no branch of low frequency oscillations in a non-rotating barotropic planet. When the planet 
rotates the peaks  corresponding to fundamental modes  are shifted leftwards, with the magnitude  of the  shift
proportional to $\Omega $. This effect is  briefly discussed  below. 

The most notable difference between the non-rotating case and the rotating ones is the presence of peaks in between
the two peaks corresponding to the fundamental modes in the rotating cases. These are the peaks corresponding to
the inertial modes. Note that all these peaks lie within the theoretical range  allowed for  inertial waves 
$|\sigma \le 2\Omega|.$ This range  is marked by vertical lines  with  styles corresponding to those   
associated with the corresponding angular velocity.
 For  small values of $\Omega $, significantly less than the largest
value $\Omega/\Omega_* =0.6$ shown in  Fig. \ref{fn2} , there are two notably large peaks within the inertial
range, one with a positive and one with a negative value of $\sigma $. 
These are identified with two main global modes discussed in detail  in PI, IP and IPN. From the point of view of
the WKBJ theory developed in IPN, these modes correspond to smallest value of an integer $l$ classifying
WKBJ modes, called the WKBJ order later on.  In the case of $\Omega/\Omega_*=0.6,$ represented by 
the dot-dashed curve, there is a third noticeable peak close to $\sigma \approx 2\Omega $  and within the inertial mode
range. It is the prograde fundamental mode, which has been  shifted within this frequency 
 range by rotation. A number of peaks
with smaller amplitudes give the positions of higher order inertial modes.
\begin{figure}
\vspace{2cm}
\epsfig{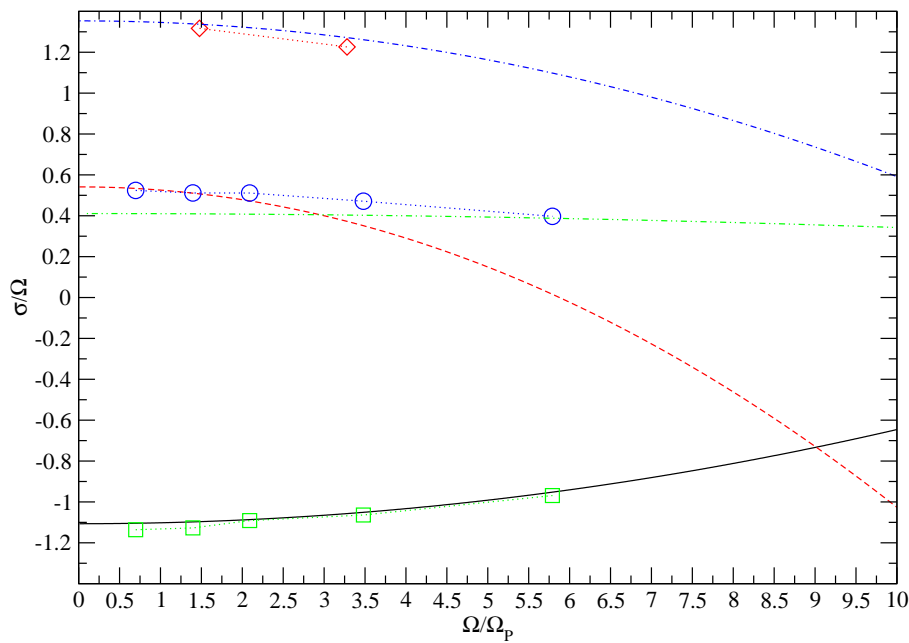}\vspace{1cm}\caption{ 
Eigenfrequencies $\sigma /\Omega $ 
of a few 'global' inertial modes as functions of the ratio
$\Omega/\Omega_{p}$, where $\Omega_{p}$ is is a typical frequency of
periastron passage, see equation (\ref{a1}).
The dot-dashed curve having $\sigma /\Omega \approx 1.3$ when $\Omega\rightarrow 0$ is the analytical 
curve for the axisymmetric $m=0$ global $l=2$ mode. The curves having $\sigma/\Omega \approx 0.5$ and
$-1.1$ when $\Omega\rightarrow 0$ are for the global $l=1$ $m=2$ modes, while the curve with  
$\sigma/\Omega \approx 0.4$ is for the next order $l=2$ curve. Symbols joined by dotted lines represent 
the corresponding results of numerical calculations. One can see that the agreement between the numerical 
and analytical results for the global $m=0$ mode and the global $m=2$ mode with negative values of 
$\sigma $ is quite good. The analytical 
results for the global $m=2$ mode with positive values of $\sigma $ agree with the numerical ones only
when $\Omega $ is sufficiently small. In the opposite case of large values of $\Omega $ the numerical 
results are close to the analytical curve for the next order $m=2$ $l=2$ mode. This result may be explained
by the effect of avoided crossing.
\label{fn5}}
\end{figure} 


\begin{figure}
\vspace{2cm}
\epsfig{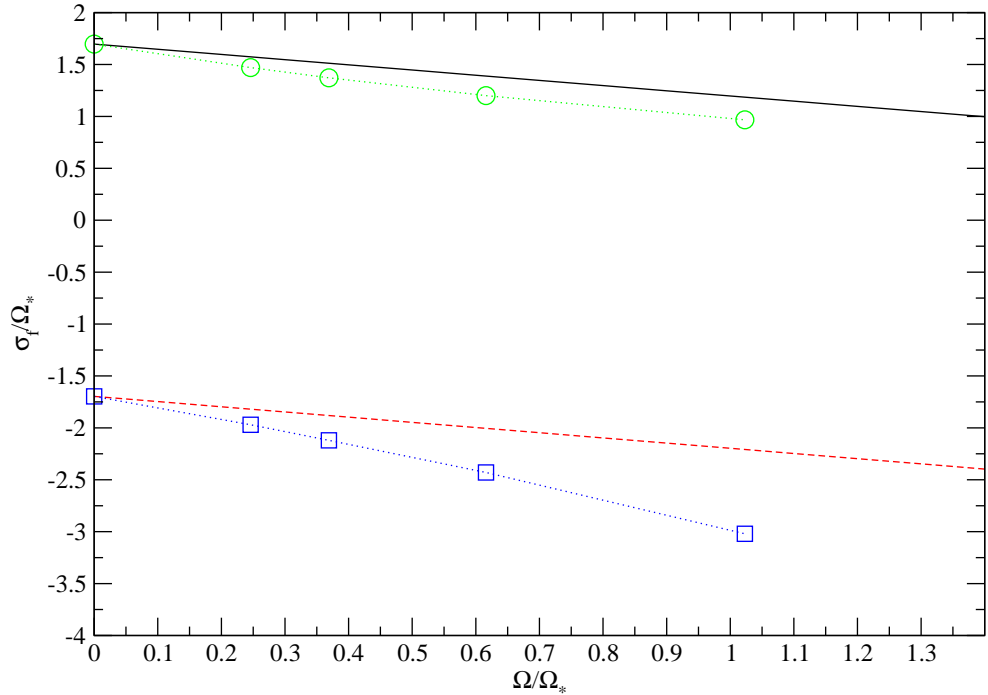}\vspace{1cm}\caption{ 
Eigenfrequencies $\sigma /\Omega_{*} $ 
of $m=2$ fundamental modes as functions of the ratio $\Omega/\Omega_{*}$. The solid and dashed curves
represent the theoretical results for the prograde and retrograde mode, respectively. Symbols joined 
by the dotted lines represent the results of numerical calculation. One can see that the agreement between
the theoretical and numerical results is much better for the prograde mode. Note that the same phenomenon 
was observed in numerical calculations of normal modes of rotating polytropic stars, which do not use  
the Cowling approximation, ( see eg. Managan 1986).
\label{fn6}}
\end{figure} 
The eigenfrequencies of inertial modes are naturally expressed in units of $\Omega$ and 
 we show the part of the spectrum  within the inertial range as a function of the ratio $\sigma/\Omega$, 
for the case $m=2$ in Fig. \ref{fn3} and for the case $m=0$ in Fig. \ref{fn4}.  Additionally, we show 
positions of WKBJ modes calculated in IPN by symbols with larger values of symbol ordinates and sizes 
corresponding to smaller values of the WKBJ order $l$. As was discussed in IPN among the WKBJ modes
with the same value of $l$ there are modes, which were not identified with results obtained by other 
methods. These modes are situated close to boundaries of the allowed region $\sigma=\pm 2\Omega $ as well
as close to the origin $\sigma=0$. The theory developed in IPN is
likely to be beyond limit of its applicability close to the boundaries
and to the origin $\sigma=0$. We expect that these unidentified modes
could be absent in a more advanced WKBJ theory. Therefore, we do not
show positions of modes having most negative, most positive and the
smallest values of $\sigma $,  for a given value of $l$.

In Fig. \ref{fn3} the case of $m=2$ is shown. The solid black and dashed red  curves correspond to the cases of relatively small
$\Omega/\Omega_*=0.24$ and $0.36,$ which should be compared with WKBJ results obtained in the anelastic 
approximation. Two global modes are clearly seen with their positions in a close agreement with positions
of the WKBJ modes having the smallest value of $l_{min}=1$ for $m=2.$
 The WKBJ modes corresponding to 
 $l=2$ and $3$ are also shown. Their positions are close to some numerically 
obtained peaks, notably in the case of the  $l=2$ modes having the ratio $\sigma/\Omega $ close to $-1.5$, 
$-0.8$ and $1.25$. The fact that some of the next order WKBJ modes are not seen in the numerical data 
is not  worrying since the amplitude of modes excited after the tidal encounter depends of the shape 
of the Fourier transform of the dependence of the tidal potential on time, which, in its turn, depends
very sensitively on the value of $\Omega,$  see PI and IP.  With increase of $\Omega $ 
the retrograde modes with negative values of $\sigma $ are excited more efficiently and their amplitude
grows, see Figure \ref{fn3}. In the case of the largest considered value of $\Omega=\Omega_*$
the peaks corresponding to the retrograde mode with $l=1$ and $2$ are strongly amplified, and the peaks
close to $\sigma/\Omega = -1.7$ and $-0.7$ may be identified with the WKBJ modes having $l=3$. Also, the 
peaks are shifted towards the origin $\sigma=0$ as  $\Omega $ increases, 
 an  effect  which may be explained by
corrections to the anelastic approximation, see IPN and  below.
 In the cases of $\Omega/\Omega_*= 0.6$ (the dotted  green curve) and $\Omega/\Omega_*= 1$ (the dot-dashed blue curve)
there is an additional noticeable peak with a positive value of $\sigma$.
This corresponds to the prograde fundamental mode being shifted into the inertial range due to rotation.              
\ \ \ \newline
\indent In Fig. \ref{fn4} we show the case $m=0$. Since in this case the spectra are symmetric about $\sigma=0$ only
positive frequencies within  the inertial range $\sigma < 2\Omega $ are considered. Similar to the previous case
symbols show positions of the WKBJ eigenmodes calculated by IPN.  
 As in Fig. \ref{fn2}  larger ordinates correspond to smaller values of the WKBJ order $l$ 
with the smallest  $l=l_{min}=0$ denoted  by circles and the eigenfrequencies of 
modes with largest shown $l=4$ denoted by stars. We see from Fig.\ref{fn4} that the peak with largest amplitude
has a value of $\sigma$ close to the identified main global mode with $l=0$ and $\sigma \approx 1.3\Omega $.
Two  identified WKBJ modes at the next order,
  also have well pronounced peaks with positions close to the values
of their eigenfrequencies. One can also possibly identify the eigenfrequencies corresponding 
higher  order modes with some peaks with smaller amplitude, especially   in the region $0.5 < 
\sigma/\Omega < 1$, where there is seemingly a good agreement between positions of the peaks and eigenfrequencies
of modes having $l\le 6$. When rotation of the planet increases the 
positions of the peaks are shifted to smaller
values of $\sigma $. This agrees with results obtained in IPN, see also below.        

\subsubsection{Change of eigenfrequencies with $\Omega $}
In the anelastic approximation  which is frequently used for calculations of eigenspectra of modes belonging
to the inertial branch,  the mode frequencies are proportional to $\Omega $. This approximation
is not, however, exact, and it discards certain terms proportional to $(\Omega/\Omega_{*})^{2}$ in 
the set of equations describing pulsations, see eg. IPN for discussion. In IPN a correction to the mode 
eigenfrequencies due to terms not accounted for in the anelastic approximation was calculated. It was shown
that in the linear approximation in the small parameter  $(\Omega/\Omega_{*})^{2}$
and neglecting a possible mode mixing, the eigenfrequencies have the form
\begin{equation}
\sigma_{i}=\sigma^{a}_{i}+\alpha_{i}(\Omega/\Omega_{*})^{2}\Omega,
\label{ne2}
\end{equation}        
where $\sigma_{i}^{a}$ is the  value of a certain eigenfrequency calculated in the anelastic approximation and 
an explicit expression for dimensionless parameter $\alpha_{i}$ can be found in IPN. We calculate $\alpha_{i}$
for three modes corresponding to the case $m=2$ using results obtained in PI and IP. For the main retrograde
global mode having $l=l_{min}=1,$ from the viewpoint of the WKBJ method, the eigenfrequency $\sigma^{a}_{1}\approx 
-1.1\Omega$ and $\alpha_{1}\approx 0.15$. For the prograde $l=1$ main global mode we have $\sigma^{a}_{2}\approx 
0.54\Omega$ and $\alpha_{2}\approx -0.5$, and for the next order $l=2$ mode which has a close value of the  eigenfrequency 
$\sigma_{3}\approx 0.41\Omega$ we get $\alpha_{3}\approx -0.022$. Additionally, we calculate the correction for 
the $m=2$ main global mode with $l=l_{min}=2$, which has $\sigma_{1}\approx 1.35\Omega $ and 
$\alpha_{1}\approx -0.12$. Then, the value of $\sigma_{i}$ given by equation (\ref{ne2}) is compared with
what is obtained in the numerical calculations. The results of this comparison are presented in Figure 
\ref{fn5}. Note that for the curves corresponding to $m=2$ the parameter $\eta=4\sqrt{2}$ while in the case $m=0,$
$\eta=4$, and we recall that $\eta=\Omega_{*}/\Omega_{p}$, where
$\Omega_{p}$ is is a typical frequency of  periastron passage
explicitly defined in
equation (\ref{a1}) of appendix A. As seen from  Fig.\ref{fn5}  there is  
good agreement between the numerical results and those calculated 
with help of equation (\ref{ne2}) for the $m=0$ global mode and 
the retrograde $m=2$ global mode. In the case of the prograde $m=2$ global mode only two numerical points
with $\Omega/\Omega_{p} < 1.5$ are in agreement with the analytical result. For larger rotation rates 
the numerical curve deviates from the analytical one calculated for the global mode. Instead, it approaches
the analytical curve calculated for the next order $l=2$ mode in the limit of large rotation. Since the
theoretical curves corresponding to the global mode and the $l=2$ mode intersect each other at 
$\Omega/\Omega_{p} \approx 3$, the fact that at larger rotation rate the numerical results are closer to
the curve describing the $l=2$ mode may be explained by the phenomenon of avoided crossing. The global and
the next order mode lose their identity at these rotation rates, being in mixed states with mode mixing
provided by an operator describing the correction to the anelastic approximation, 
see IPN for its explicit form.

Now let us shortly discuss the behaviour of the fundamental modes with change of rotation rate. In order to 
find the corresponding analytical estimate of change of eigenfrequencies we should remember that we work
in the rotating frame while the fundamental modes are normally studied in the inertial frame. In the latter
frame there is a simple expression for a correction of the mode eigenfrequency, $\sigma_{f}$, due to rotation
\begin{equation}
\sigma_{f}-\sigma^{nr}_{f}=m\beta \Omega,
\label{ne3} 
\end{equation}     
where $\sigma_{f}^{nr}$ is the frequency of a fundamental mode calculated for a non-rotating planet, and
$\beta $ is determined by some integral over the planet's volume, see eg. Cristensen-Dalsgaard and 
references therein. As was shown by eq. Ivanov $\& $ Papaloizou 2004 $\beta \approx 0.5$ for planetary 
models, which are close to the standard $n=1$ polytrope. Eigenfrequencies associated with the rotating frame
are shifted with respect to the ones in the inertial frame according to the rule $\sigma \rightarrow 
\sigma - m\Omega$. Therefore, in the rotating frame, for the case of the  $m=2$ mode, we obtain a simple
relation
\begin{equation}
\sigma_{f}=\sigma^{nr}_{f}-\Omega,
\label{ne4} 
\end{equation}   
where we set $\beta=0.5$. This relation is compared with the numerical results in Fig. \ref{fn6}. One 
can see from  Fig.\ref{fn6}  that the numerical and analytical approaches are in a good agreement for the 
prograde mode while in the case of retrograde mode the results differ. The fact that the simple approximation
(\ref{ne4}) doesn't work well for the retrograde fundamental modes, for finite values of rotation rates,
has been known for quite a long time, see eg. Managan (1986). Since this issue is not directly relevant to the 
purposes of this paper we do not discuss it in more detail here. 
  
  \section{ Coreless models under a  fixed quadratic potential }

Goodman \& Lackner (2009)  considered special barotropic models
which are in hydrostatic equilibrium  under a fixed
quadratic gravitational potential $\Psi$ in the unperturbed state.
Thus
 \begin{equation}
   \frac{1}{\rho}\frac{dP}{dr} =  \frac{c^2_s}{\rho}\frac{d\rho}{dr}=
-\frac{d\Psi}{dr},
   \end{equation}
with $\Psi = c_1+c_2r^2,$ with $c_1$ and $c_2$ being constant.
Such models have a special property in common with the uniform density
model that $(1/\varpi)\partial \Psi/\partial \varpi
= (1/r)d\Psi/dr=c_2 \equiv \omega_0^2$ is constant.
Such spherically symmetric models may be found for an arbitrary  specified
density distribution if the pressure is allowed to be determined {\it a posteriori}
by the condition for hydrostatic equilibrium.

Goodman \& Lackner (2009) showed that  these models had the property
that when full compressibility was retained,  no inertial modes are excited by
a quadrupole $m=2$ tidal potential. Thus for these models the response
is determined entirely by the $f$ modes. In appendix \ref{Goodman models}
we review the compressible case and show that  no inertial modes
are excited when the anelastic approximation we use is employed.
Thus no spurious inertial mode excitation occurs as a result of its use for these models.
Nonetheless it is important that the numerical schemes we employ  correctly 
represent the tidal excitation of these models. This we demonstrate below
for the finite difference approach adopted in this paper. The situation with respect 
to the spectral approach adopted in previous papers, which is also found to behave correctly
is discussed in appendix \ref{Goodman models}.

 \subsection{Simulations with quadratic potential models}
  The models that we  have adopted 
  in order to compare their tidal response to that
  of  the standard polytropic model considered in this paper  have the same mass and radius as
  but a density distribution that
  depends quadratically on radius.
  The central density is given by 
   \begin{equation}
  \rho_c=\frac{15M_*}{8\pi R_*^3}
  \end{equation}
  This is smaller than the corresponding value
  for the polytrope in equilibrium  under a self-consistent gravitational potential 
  given in  appendix  \ref{appenA} because the
  quadratic potential model is less centrally condensed.
  The central pressure is given by
     \begin{equation}
      P_c=\frac{15 G M_*^2}{32\pi R_*^4}.
    \end{equation}
The density and pressure in the interior are then respectively  given by
  $\rho  =\rho_c(1-r^2/R_*^2)$and
  $P  =  P_c(1-r^2/R_*^2)^2.$
  Thus this model also satisfies the equation of state for a polytrope with
 index  $n=1.$
However, hydrostatic equilibrium    confirms that this is a quadratic
potential model with potential determined to within an
arbitrary constant by
$\Psi = GM_*r^2/(2R_*^3).$

\begin{figure}
\vspace{1cm}
\epsfig{file=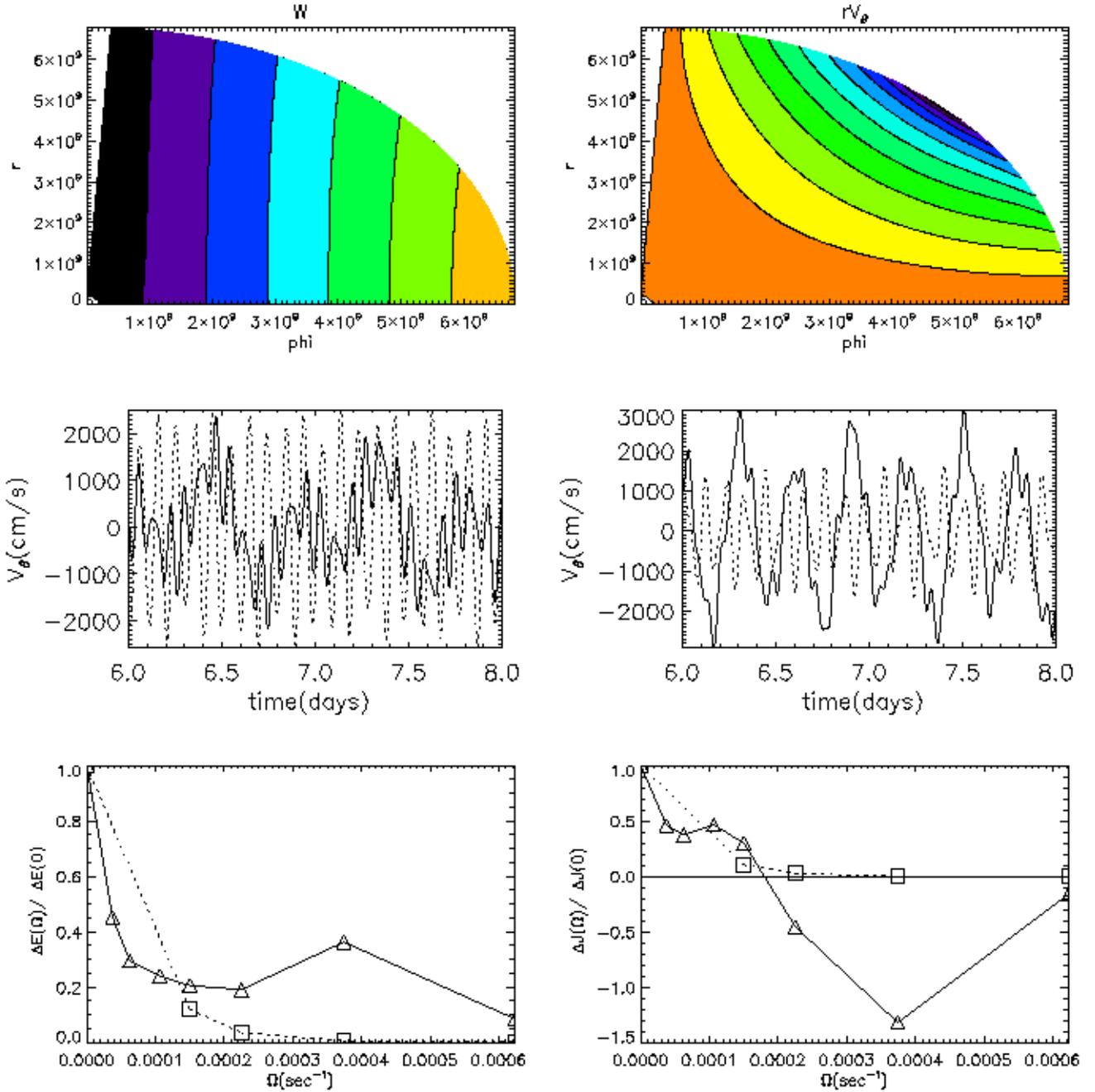}\vspace{1cm}\caption{This figure illustrates
the behaviour of the   quadratic potential models  discussed
in the text  with  some comparisons
to a corresponding standard polytropic model for encounters with $\eta=4\sqrt{2}.$
The left  uppermost  panel gives contours of the real part of  $W$ and the 
right uppermost panel contours of the real part of $rv_{\theta}$ 
 $13.38$ days after pericentre
passage for the quadratic potential  model with
 $\Omega/\Omega_*=0.36.$
The   central panels give the real part of $v_{\theta}$ 
evaluated at $r=4.8\times10^9cm$ and $\theta= 67$ degrees
as a function of time
between six and eight days after pericentre passage. The left central panel
illustrates the standard polytropic model (solid curve) and the corresponding
quadratic potential model (dashed curve) for 
 $\Omega/\Omega_*= 0.24.$
The right central panel illustrates the same 
quantities for $\Omega/\Omega_*= 0.36.$
The lowermost left panel shows the energy transferred from the orbit relative to its value for $\Omega =0$ 
as a function of $\Omega$ in sec$^{-1}.$ {  We recall that $\Omega_*=6.23\times 10^{-4}$~sec~$^{-1}.$}  
 The standard polytrope is illustrated with the solid  curve and the quadratic potential
model with the dotted curve. The lowermost right panel gives the corresponding plot for
the angular momentum transferred from the orbit.}
\label{Goodmanfig}
\end{figure}

\begin{figure}
\vspace{2cm}
\epsfig{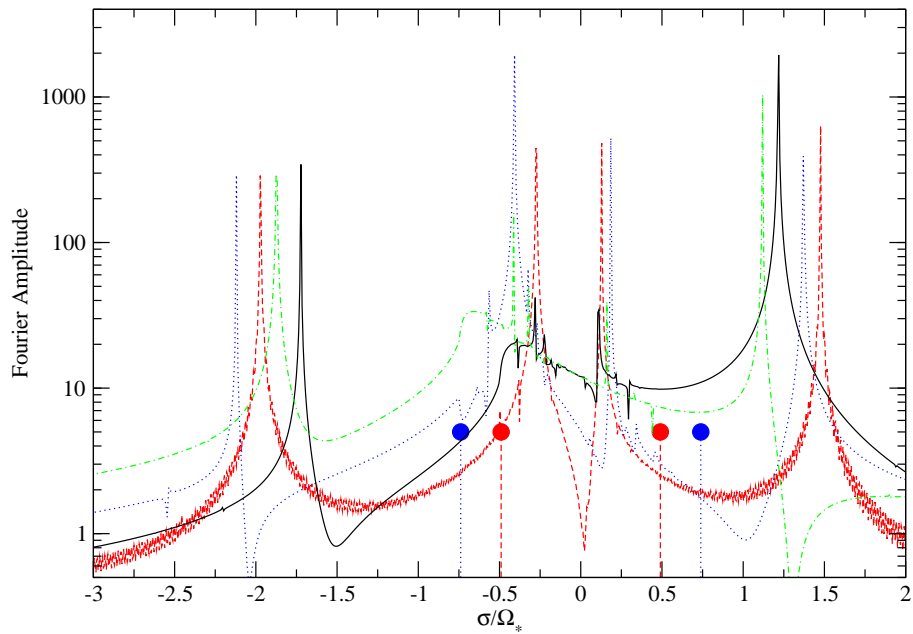}\vspace{1cm}\caption{ This shows a  comparison of the 
Fourier spectra obtained by the Fourier transform of the velocity data as functions of the
ratio $\sigma/\Omega_{*}$ for  $\eta=4\sqrt{2}$ for  standard polytropic and corresponding
quadratic potential models.
 The case $m=2$ is shown.
The solid (black) curve corresponds to the quadratic potential model   with $\Omega/\Omega_*= 0.24$ while
the dashed  (red) curve is for the corresponding polytropic model.
The  dash-dotted (green) curve is for the quadratic potential model with
 $\Omega/\Omega_*= 0.36$ and the dotted  (blue) curve is for the corresponding polytropic model.
The vertical lines delimit the range of
$\sigma $, where  inertial modes are  theoretically possible, with
 line styles identical to those used for
corresponding
values of $\Omega.$}
\label{Goodmanspectra}
\end{figure}

{\begin{table}
\begin{center}
  \begin{tabular}{|c|c|c|c|c|}
    \hline
   & $ f_{+a}$ & $ f_{+n}$ & $f_{-a}$ & $f_{-n}$ \\ \hline
 $\Omega /\Omega_*$ & $\sigma/\Omega_*$ & $\sigma/\Omega_*$  & $\sigma/\Omega_*$ & $\sigma/\Omega_* $ \\ \hline
$ 0$& 1.4142 & 1.4154 &-1.4142 &-1.4154 \\ \hline
$0.24$& 1.1940 & 1.1965 &-1.6750 &-1.6751 \\  \hline
$0.36$ & 1.0988 & 1.0990 &-1.8202  &-1.8222 \\
    \hline
  \end{tabular}
\caption{A comparison between $f$ mode angular  frequencies obtained
from numerical simulations with analytically determined values.
The first column gives the angular velocity, the second and third columns
give the analytic and numerical prograde $f$  mode angular frequencies respectively. The fourth and fifth columns give the corresponding quantities for
the retrograde $f$ mode.
\label{Goodmantable}}
\end{center}
\end{table}}

   We  have performed simulations   of quadratic potential models for
  $\eta =4\sqrt{2}$  for a number of rotation frequencies and a resolution $200\times200$
   that can be directly compared
  to those for the corresponding standard polytropes with $n=1$ (the global simulation
  parameters for the different models are the same).
  In particular, we establish that although there is some mild distortion
  due to the form of the computational grid and numerical truncation, 
  that causes some weak inertial mode excitation,
  the response of the quadratic potential models is determined by the $f$ modes as expected.
  Some numerical results for these simulations are shown in Fig. \ref{Goodmanfig}.

The lowermost left panel shows the energy transferred from the orbit 
as a function of $\Omega$ relative to its value for $\Omega =0.$
The lowermost right panel gives the corresponding plot for
the angular momentum transferred from the orbit.
We remark that when $\Omega=0,$ the energy and angular momentum transfers
for the quadratic potential models exceed those for the polytropic models
by factors of $9.2 $ and $11.0 $ respectively,  because the polytropic models are more centrally condensed.
The lack of inertial mode response is apparent in the quadratic potential
models which do not show the related increase in energy transfer
for intermediate values of $\Omega.$ The interaction is much weaker
for intermediate and larger values of $\Omega$ for these models.
This is also indicated when one considers the value of $\Omega$ for which
the angular momentum transfer is zero. This defines the angular velocity
$\Omega_{ps}$ for which pseudo synchronisation is achieved (Ivanov \& Papaloizou 2004, IP). 

   The angular velocity for pseudo synchronisation 
   taking account only $f$ modes was estimated  in general by
   Ivanov \& Papaloizou (2004) to be given  by
   \begin{equation}
   \Omega_{ps}=\frac{(2+\ln(\eta) )\Omega_*}{\eta}.
   \end{equation}
   However, when inertial modes are taken into account IP
   find that $\Omega_{ps} \sim 1.55\Omega_*/ \eta.$
   Thus, assuming $f$ modes dominate the
   quadratic potential models and inertial modes dominate the polytropic models,
    for the simulations illustrated  in Fig. \ref {Goodmanfig}, we expect
   $\Omega_{ps} = 071\Omega_*$ in the former case
   and  $\Omega_{ps} = 0.27\Omega_*$ in the latter.
   For the numerical results  plotted in Fig. \ref {Goodmanfig},
    $\Omega_{ps} \sim  0.6\Omega_*$ 
   and  $\Omega_{ps}  \sim  0.29\Omega_*$ for these 
   quantities respectively. As  the  estimates of Ivanov \& Papaloizou (2004)
   and IP assumed $\Omega$ to be small, this agreement is satisfactory
    and implies consistency with the view that the polytropic models
     are inertial mode dominated and the quadratic potential models $f$ mode dominated.

  The form of 
  the tidally excited disturbance in the quadratic potential model
  with  $\Omega/\Omega_*= 0.36$
 $13.38$ days after pericentre is illustrated in the uppermost panels of Fig. \ref{Goodmanfig}.
 The same form is seen at other times long enough after the tidal encounter.
Contours of the real parts of  $W$ and  $rv_{\theta}$ 
are shown. The former are approximately vertical as expected
from the analytic form $W\propto \varpi^2$ (see appendix \ref{Goodman models})  and the latter
are approximately hyperbolae corresponding to the
expected form $rv_{\theta}\propto \varpi z.$    

 The   central panels of Fig. \ref{Goodmanfig} give  the real part of $v_{\theta}$ as a function of time
at an inner point of the planet,  with $r=4.8\times 10^9 cm$
and $\theta = 67$ degrees,
between six and eight days after pericentre passage both for
 the standard  polytropic model  and the corresponding
quadratic potential model for 
 $\Omega/\Omega_*=0.24$
and $\Omega/\Omega_*=0.36.$
It will be seen that the quadratic potential model responses  show high frequency
behaviour corresponding to excitation of the $f$ modes whereas the corresponding polytropic
models show in addition a dominant contribution from longer period disturbances that are  associated with
inertial modes.
  
In order to better quantify this  we have performed  a  Fourier analysis of  longer spans
of these time series in order to determine the mode spectra for these cases.
The results are plotted in Fig. \ref{Goodmanspectra}. From this figure it is apparent that
when corresponding models are compared the inertial mode amplitudes are an order of magnitude
or more smaller for the quadratic potential models, corresponding to a mode energy
a hundred times or more smaller.  For the standard polytropic cases plotted, the
response is dominated by the inertial modes in contrast to the $f$ mode dominated 
quadratic potential models. 

As an additional test we compare the numerically determined $f$ mode frequencies
for the quadratic potential models to the analytically determined values.
For the quadratic potential models the $f$ mode angular frequencies are given by
$\sigma = -\Omega \pm\sqrt{2\omega_0^2+\Omega^2}$ (Goodman \& Lackner 2009).
Here the positive sign corresponds to the prograde mode and the negative 
sign to the retrograde mode.
We compare these values to the numerical values we obtained
from the spectral analysis of the time series indicated above
for  $\Omega=0, \Omega/\Omega_* = 0.24$ 
   and  $\Omega/\Omega_*=0.36$ in table \ref{Goodmantable}.
   The  numerical  and  analytic values correspond to within  a relative discrepancy of $2\times 10^{-3}$ 
   and so are in good agreement.   We conclude that our numerical treatment
   adequately represents the $f$ modes and their dominance of the tidal response 
   for the quadratic potential models.


\begin{figure}
\vspace{2cm}
\epsfig{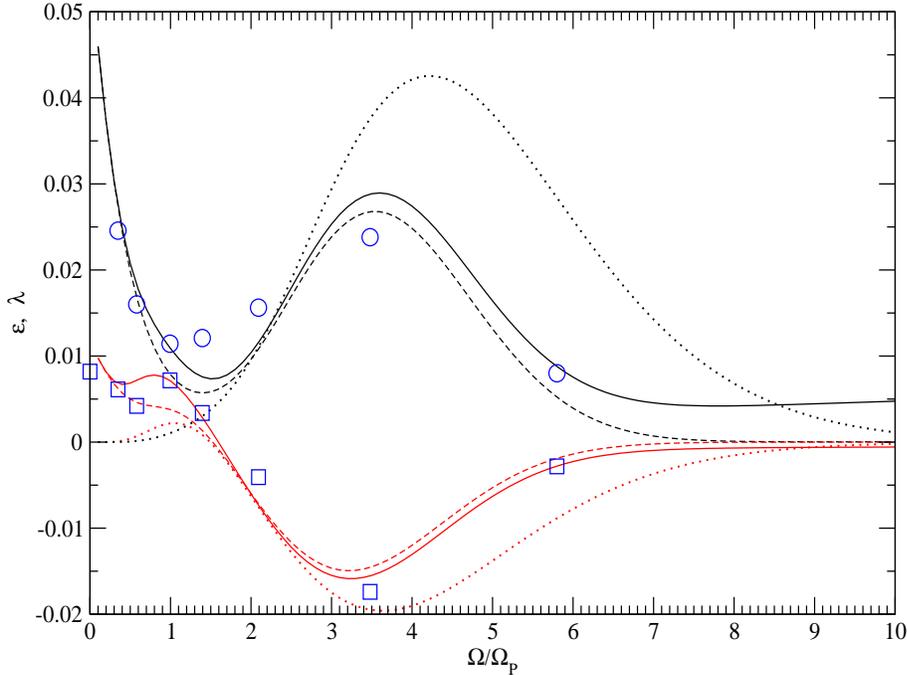}\vspace{1cm}\caption{ 
The dimensionless quantities $\epsilon $ and $\lambda $ determining the energy transfer in the rotating frame and 
the angular momentum transfer, respectively, are plotted as functions of 
$\Omega/\Omega_{p}$. The case of a tidal encounter having
$\eta=4\sqrt{2}$ is shown. The uppermost solid, dashed
and dotted curves describe results of theoretical calculation of $\epsilon$ while the corresponding 
lowermost curves determine theoretical values of $\lambda $. Symbols represent the results of numerical 
calculations. The solid curves are calculated with all inertial and fundamental modes accounted for,
the frequency corrections to the fundamental modes given by equation
(\ref{ne4}) and to two main global modes given by equation (\ref{ne2}) are taken into account 
as well. The dashed curve differ from the solid ones by the fact that
only the two main global modes are taken 
into account in the expressions determining the contribution of the inertial modes. 
The dotted curves show the case when the contribution of the fundamental modes as well as
the frequency corrections are neglected and only two main global inertial modes are taken into account. 
Note that the dotted curves do not depend on value of $\eta$. These
curves are, therefore,  called 'the universal curves'
later on.
\label{fn7}}
\end{figure} 



\begin{figure}
\vspace{2cm}
\epsfig{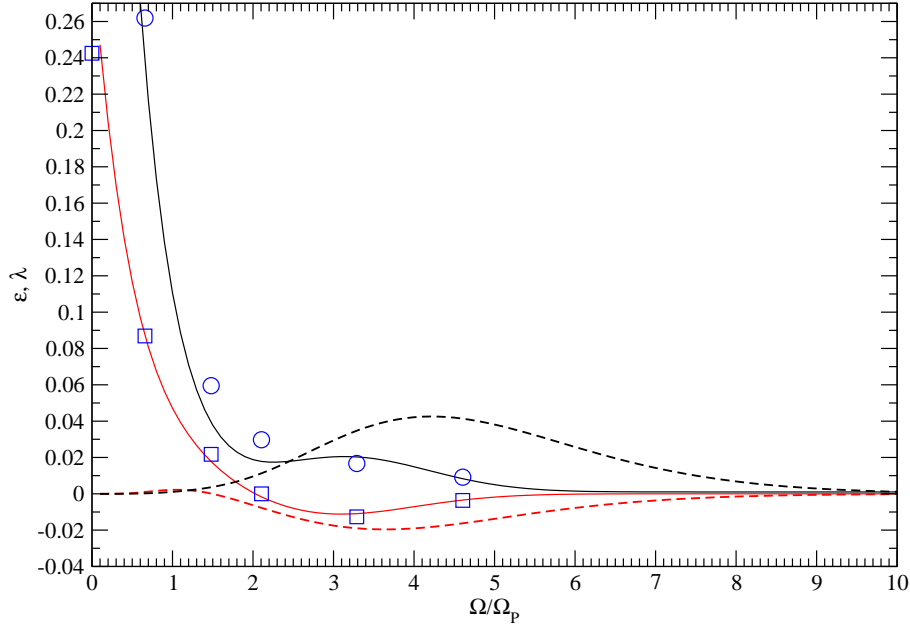}\vspace{1cm}\caption{ 
Same as Fig. \ref{fn7} but for   a closer encounter with $\eta=4$. For
simplicity only  two mail global modes are taken into account in the
expressions determining the energy and angular momentum transfer due
to inertial waves. Both contributions of the fundamental modes and of
inertial modes as well as the corresponding frequency corrections are
used to calculate the solid curves while the dashed curves are the
universal ones.
\label{fn9}}
\end{figure} 
\begin{figure}
\vspace{2cm}
\epsfig{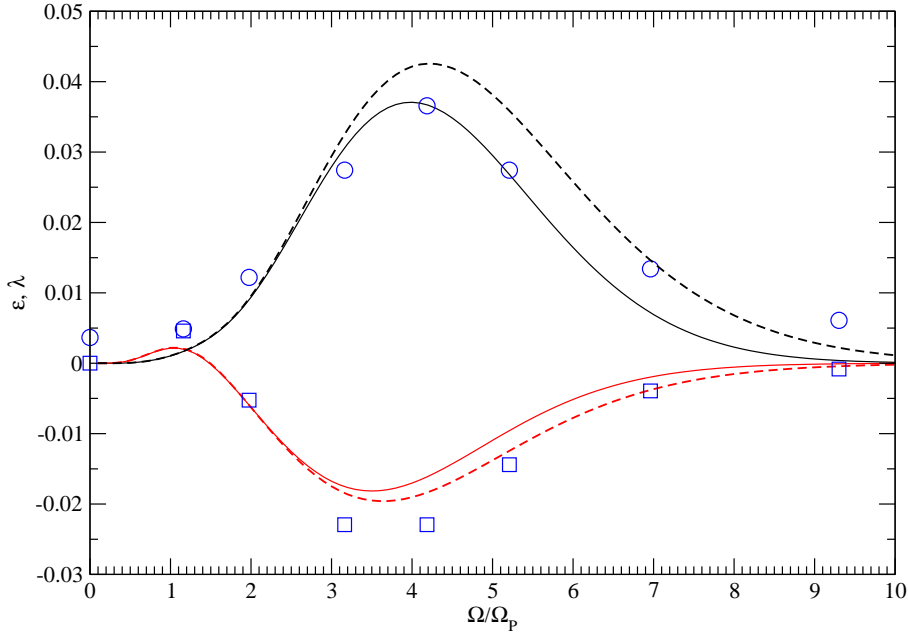}\vspace{1cm}\caption{ 
Same as Fig. \ref{fn9} but for  a more distant encounter with  $\eta=8\sqrt {2}$.
\label{fn10}}
\end{figure} 

\section{Comparison of the energy and angular momentum transfer with the results obtained
by the spectral methods}

\subsection{Numerical  simulations  versus the spectral approach}

Let us compare the energy and angular momentum
transferred for individual simulations with  results  obtained
from the  basis function (or spectral) approach developed in PI and IP.
The expressions for energy and angular momentum transfer determined by
the inertial modes, $\Delta E_{m}$ and $\Delta L_{2}$ obtained by PI in the anelastic
approximation are for convenience reproduced in appendix B, see equations (\ref{eq 26}) and
(\ref{eq 27}). Let us recall that $m$ is the azimuthal  mode number. As we
discussed above the numerical results show that modes with $m=0$ give
a negligible contribution to the energy transfer, and, therefore, we set
$m=2$ later on. Also note that the quantity $\Delta
E_{2}$ is defined in the rotating frame. The energy transfer in the
inertial frame is given as $\Delta E_{I}=\Delta E_{2}+\Omega \Delta
L_{2}$. Contrary to  $\Delta
E_{2}$, $\Delta E_{I}$ can be negative.

It follows from equations (\ref{eq 26}) and (\ref{eq 27}) that 
dimensionless quantities
\begin{equation} 
\epsilon=\eta^{6}\Delta E_{2}/E_{*}\quad \lambda= \eta^{5} \Delta
L_{2}/L_{*}
\label{ne5}
\end{equation} 
depend only on the dimensionless ratio $\Omega/\Omega_{p}$ in the
anelastic approximation.  Therefore, they conveniently represent the
energy and momentum transferred and we compare these quantities with
what is given by numerical calculations below. In our spectral
approach we use the results obtained by IP to calculate the
overlap integrals (see equation  (\ref{eq 27a}) for the  definition) and
eigenfrequencies, for their model of a cold Jupiter  with smoothed
density jump in a place of phase transition 
between molecular and metallic hydrogen, as
this model is like  a standard polytrope with  index $n=1.$ 
Formally, the quantities
$\epsilon $ and $\lambda $ are given by infinite summation series over all
eigenmodes with $m=2$. In practice, the number of modes is limited by
the resolution of the numerical method used , which in turn is
determined by the number of basis functions in  the case of a spectral
method. As we mentioned above, in the calculations of PI and IP $225$ basic functions were used
giving $450$ eigenfrequencies. However, as was mentioned by PI and IP
only two main global modes (a  prograde one with $\sigma \approx
0.5\Omega $ and a  retrograde one with   $\sigma \approx
-\Omega $ give the  main contribution to the transfer of energy and
angular momentum, for small and moderate ratios of
$\Omega/\Omega_{p}$. Thus, to compare the  results obtained  by different
methods we consider  two approaches:  the first  takes into account 
all  of the  eigenmodes available in the spectral method, the second
takes into account only the contributions of the 
two main global modes.

In order to compare the spectral and numerical results we should add
to the expressions (\ref{ne5}) the corresponding contributions  arising from 
excitation of the fundamental modes, for a rotating planet. These have 
been derived in Ivanov $\&$ Papaloizou 2004 (see also Lai
1997)\footnote{Note that the final expressions of  Ivanov $\&$
Papaloizou 2004 should be multiplied by $\pi^{2}$ due to a misprint in
the text, the correct expressions are given in an appendix of
IP.}. Additionally, we take into account the frequency correction
(\ref{ne2}). After these modifications  are made, the expressions
(\ref{ne5}) acquire dependence on $\eta $. To emphasise the difference
between the expressions with and without these modifications we call
the latter  'universal curves'.

The results of the comparison of $\epsilon $ and $\lambda $ are
presented in Figures \ref{fn7}-\ref{fn10}.  
In Fig. \ref{fn7} we show the case of a tidal encounter with
$\eta=4\sqrt 2$. The solid and dashed curves represent the results
given by the spectral method with the contribution of the fundamental 
modes as well as the frequency corrections included while the dotted
curves are the universal ones.  The symbols show the results of
numerical calculations. One can see that the agreement between the 
spectral and analytical approaches is quite good apart, possibly a
region with $\Omega/\Omega_{p}\approx 1.5$. In particular, it is clear
from  Fig.\ref{fn7}  that for a sufficiently small ratio $\Omega/\Omega_{p} <
1.5$ the energy and angular momentum gained by the planet are
determined by excitation of the fundamental modes while in the
opposite case the inertial waves dominate the tidal
response. Therefore, eg. the curve giving the energy transferred to the
planet has non-monotonic shape. Initially it decreases with 
$\Omega/\Omega_{p}$ due to suppression of the contribution determined by the fundamental modes
caused by increase of rotation, then it starts to increase due to
increasingly more efficient excitation of the inertial waves attaining
a maximum at $\Omega/\Omega_{p}\sim 3.5$, then it decreases again. One
can also see that the role played by the frequency corrections to the
inertial modes is essential with the universal curves giving much
worse agreement with the numerical results in the region, where the
inertial waves dominate. On the other hand it is  difficult to see whether 
the curve with all inertial eigenmodes taken into account (the solid
curve) or the curve determined by two main global modes (the dashed
curve) gives a better approximation to the numerical data. We expect
that in order to differentiate between these two curves even larger 
rotation rates should be considered. This issue is briefly 
discussed below, where we compare results of the
spectral and WKBJ methods. In the rest of this section we consider 
only two main global modes in the expressions for the tidal transfer 
due to inertial waves. 

In Fig. (\ref{fn9}) the case of a stronger tidal encounter with
smaller value of $\eta=4$ is considered. As in the previous case the
agreement is very good. As expected the role of the
fundamental modes is more important at small rotation rate while the
role of the frequency corrections becomes quite important at large 
values of rotation. 

Fig. (\ref{fn10}) shows the case of relatively weak tidal encounter
with $\eta=8\sqrt 2$. In this case the inertial waves dominates the
tidal response for all non-zero rotation rates considered. Also, the 
role played by the frequency corrections is minor relative to the
previous case, and, therefore, the curves taking these ones into
account are close to the universal curves in this case. Note that the
numerically obtained value of $\epsilon $ is larger than the
theoretical value for the maximal $\Omega/\Omega_{p}\approx 9$.  This
might possibly  be explained by contributions of higher order WKBJ modes to the
energy transfer, which are not taken into account in the theoretical
curve.

\section{Discussion}

In this paper we extended  our previous work (IP)
on the tidal interaction of a planet on  a highly eccentric 
or parabolic orbit  about a central star  by relaxing the anelastic approximation
and considering  the tidal response of planet models with solid cores.
These changes were accomplished by
carrying out  numerical simulations
of tidal encounters considered  as  initial value problems.
 The response of  $f$
and $p$ modes  was found
in addition to that due to   low frequency inertial  modes.

We calculated  the energy and angular momentum exchanged after 
 pericentre passage for a variety of pericentre distances and  angular velocities of rotation 
of the planet.  Models both  with and without a solid core were simulated.
In the latter case the state variables showed evidence of { the emission of shear layers and/or} wave attractors
after the encounter.  

In the coreless case we studied  the spectrum of excited inertial  modes 
and compared it with the spectra obtained in IP and with the analytic theory 
described in  IPN.  We found good agreement for slow rotation.
 We were also able to account 
for the variation of the normal mode frequencies with the planet's
angular velocity. In the case of inertial waves a non-trivial variation is induced by terms neglected
in the anelastic approximation and it is found to be  small  for rotation
frequencies significantly less than critical.   Thus  our analysis indicated that this
approximation is generally  a valid one for the inertial modes.

We presented results for the total energy
and angular momentum transferred to a planet without a solid core for  the full range of rotation rates and  a variety
of encounter pericentre distances and found good agreement in the coreless case
with results obtained by IP which were obtained using an anelastic approximation.  
When a solid core is introduced 
the energy  exchanged  between star and planet is  in general not changed by  a very large amount. 
We  give  both a physical and mathematical explanation of this result in appendix  B.

It is  a consequence of the good agreement
of our results with those obtained previously using other methods  that the anelastic approximation
is valid for the global inertial modes in the slow rotation limit. Thus we do not see the 
large enhancement of the tidal interaction that 
Goodman \& Lackner (2009) indicated would occur as a consequence of
adoption of the anelastic approximation.

Since in this limit  our numerical results fully agree with the previous ones obtained
in  Ivanov $\&$ Papaloizou 2004 and IP, where estimates of the
circularisation time scales of a planet on a highly eccentric orbit
have been made, we can state that these estimates are now confirmed by
 direct numerical simulations that do not use the anelastic approximation.

\section*{Acknowledgements}

We are grateful to the referee, Jeremy Goodman, for his comments, which led 
to improvement of the paper. 

PBI was supported in part by RFBR grant 08-02-00159-a, by the
governmental grant NSh-2469.2008.2 and by the Dynasty Foundation.

This paper was prepared to the press when both
P.B.I. and J. C. B. P.  took part in the Isaac Newton programme 'Dynamics of Discs
and Planets'.

\vspace{-0.4cm}

\begin{appendix}

\section{Numerical methods and set up}\label{appenA}

\subsection{Computational grid }

The simulations were performed on a $n_r \times n_{\theta}$ 
grid $(r_i,\theta_j).$ Here $n_r$ and $n_{\theta}$ are the number of gridpoints
in the radial and theta directions respectively.
After numerical experimentation we adopted the following set up
which was found to give reasonable results for the particular models
adopted that had an equation of state corresponding to a   polytrope with $n=1 .$
      
In  setting  up the radial  grid for coreless models we 
remark that setting the innermost radial grid point too close to the
origin and/or the outermost point too close to a zero pressure surface at $r=R_*$ 
results in a singular unstable behaviour. To avoid this
we set the innermost grid point to be $r_1=3.77\times10^{-2}R_*$
and the outermost point to be $r_2=0.975R_*$ and applied
the relevant boundary conditions there.  This choice of innermost point was found to 
result in regular behaviour near the origin while avoiding the real physically
singular wave propagation 
 behaviour manifested by the the models with sizable cores. Similarly
smooth behaviour was maintained close to the bounding planetary surface.
For models with solid cores $r_1$ was replaced by the core radius.
The radial grid points were then distributed  according to

\begin{equation} 
  r_i=(r^2_{1}+(i-1)h_{2})^{1/2}, \ \ \ \  i=1,2,...n_r.
\end{equation}
Here $h_{2}=(r^2_{2}-r^2_{1})/(n_r-1).$

The angular grid was defined over $0 <  \theta < \pi/2$  in terms of
$\mu_j =\cos\theta_j $ according to the 
prescription 
 \begin{equation}
\mu_j=((j-1)h_{\mu})^{2/3}, \ \ \ \   j=1,2,...n_{th},\end{equation}
where  $ h_{\mu} = 2/(2n_{th}-1).$ 
The domain $\pi/2 < \theta < \pi$ by assuming symmetry with respect
to reflection of the density distribution in the equatorial plane.
\subsubsection{Specification of the state variables on the grid}
The numerical scheme solves for the Fourier components of each state
variable defined analogously to equation (\ref{eq p1}).
These are associated with a particular azimuthal mode number, $m,$
and are complex quantities. This is taken as read throughout.
The specification of the state variables is then staggered
so that the density and pressure perturbations $\rho'$ and $P'$ 
as well as $v_{\phi}$ and $\xi_{\phi}$ are evaluated
at cell centres $(i+1/2,j+1/2).$
The velocity $v_r$ and displacement  $\xi_r$ are evaluated at $(i.j+1/2),$
and $v_{\theta}$ and $\xi_{\theta}$ are evaluated at $(i+1/2,j).$
The specification of the state variables is also staggered in time.
Thus,  $\bmth{\xi},$ $P'$ and $\rho'$ were evaluated at time
levels $k, k=1,2...$ and  ${\bf v},$  was evaluated
at the intermediate levels $k+1/2, k=1,2...$
\subsubsection{Numerical solution}
Equation (\ref{eq p3})  is solved  by use of operator splitting
by writing it in the form 
\begin{equation}
\frac{\partial {\bf  v}} {\partial t } = {\bf L}_1 +{\bf L}_2
+{\bf L}_3, \label{eq0p3}  
\end{equation} where
${\bf L}_1 =-\nabla W$, ${\bf L}_2 = -2{\bmth {\Omega}}\times {\bf v},$
and
${\bf L}_3 =  {\bf f}_{\nu}/ \rho$ and
then solving the equations
\begin{equation}
\frac{\partial {\bf  v}} {\partial t } = {\bf L}_i ,  \ \ \ \ \   i = 1,2,3 
\end{equation}
in sequence.
For the cases $i=1$ dealing with the pressure forces and external potential
a second order time symmetric explicit finite difference scheme was used.
For  $i=3$ dealing with non conservative forces, as these in general
are of the order of the truncation error,  a first order (in time)
second order in space finite difference scheme was used.
For descriptions  of similar numerical schemes used
to solve linearised equations see Papaloizou \&  Pringle (1984, 1987)
and Lin ,  Papaloizou \& Savonije (1990).
The case $i=2$ above involving Coriolis forces was  integrated 
as a straightforward system  of simultaneous    equations using a second order Runge-Kutta method.
Having advanced ${\bf v},$  $\bmath{\xi}$ was obtained
from equation (\ref {xidot}) and then to complete the cycle
$P',$ and $\rho'$ are found from
equation (\ref{eq p6}).

\subsubsection{Boundary conditions}
Boundary conditions need to be applied
at the innermost  and outermost radial grid points,
 on the equatorial  plane  $\theta=\pi/2$ and on the
 cone $\theta  = \theta_{n_{\theta}}.$
 On the inner radial boundary we adopt a regularity 
 condition or $v_r=0.$ On the outer radial
 boundary we take the Lagrangian pressure perturbation
 to be zero. Finally at both $\theta=\theta_{n_{\theta}}$ and
 $\theta= \pi/2$ we adopt $v_{\theta}=0.$

\subsubsection{Initial density}\label{A1.4}
 The initial density distribution is taken to be that of a standard polytrope
with index $n=1$  { in equilibrium under its self consistent gravitational potential.}
Thus, for all models, outside of any core  we adopt
\begin{equation}
 \rho=\frac{\rho_cR_*\sin(\pi  r/R_*)}{\pi r}\end{equation}
where the central density is given by
$\rho_c=\pi M_*/(4R_*^3)$
and $M_*$ is the total planet mass.

\subsection {The form of the tidal potential for $m=2$ and $m=0$}
The $m=2$ component of tidal potential is 
\begin{equation} \Psi_{ext,2} = -\frac{3\Omega_p^2 M}{4(M_*+M)}\frac{R_{p}^3}{R^{3}}r^2
      \sin^2(\theta)\exp( -2i \phi_p), \label{Tidpot}
\end{equation}
(see equation (\ref{eq p1}) where $M$  be  the central perturbing  mass  and $R_{p}$ be the pericentre distance.
The corresponding  $m=0$ axisymmetric  component of the tidal potential is
\begin{equation} \Psi_{ext,0} = \frac{ \Omega^2_pM}{4(M_*+M)}\frac{R_{p}^3}{R^{3}}r^2(3\mu^2-1).
\end{equation}

The radial and angular  coordinates of the planet are  given by  
\begin{equation} R = R_{p}(1+4\sinh^2x ), \  \  \ {\rm and} \ \ \ \cos(\phi_p) =2R_{p}/R-1 \end{equation}
respectively. The quantity $x$ is related to the time $t$ through
$\sinh(3x)= 3\Omega_p( t-t_0)/16,$
where 
 \begin{equation} \Omega_p= \sqrt{G(M_*+M)/R_{p}^3}\label{a1}\end{equation}
and pericentre passage occurs at $t=t_0.$
The simulations carried out in this paper started with the perturbing
mass at a distance eight times the pericentre distance. Tidal
forces being proportional to the inverse cube of the distance are negligible
beyond this point (see eg. Faber et al 2005). 
For fixed masses, as an alternative to the pericentre distance,
in this paper we define an encounter through
the  parameter
\begin{equation} \eta  =\sqrt{ M_*R_{p}^3/((M+M_*) R_*^3)} \sim \sqrt{ M_*R_{p}^3/(M R_*^3)}. \label{a2}\end {equation} 
The tidally disturbed body is explicitly taken to be a planet of one Jupiter mass
with $R_* = 7\times 10^9 cm.$

\subsection{Energy and angular momentum transfer}
The energy and angular momentum transferred during an encounter are found
by  evaluating the canonical energy and angular momentum.
These are conserved and well defined in a non dissipative system
(Friedman $\&$ Schutz 1978). In a weakly dissipative system
they decay slowly with time but their values just after the encounter
represent the associated transferred quantities.

\subsubsection{The canonical energy}

When the Lagrangian pressure perturbation vanishes
at the outer boundary, the canonical energy appropriate
to the Fourier mode with azimuthal mode number, $m,$
 may be written as
\begin{equation}
 E_c=0.25(1+\delta_{m,0})\left[ \int_V  \rho\left(   |{\bf v}|^2|    +|P'|^2/(\gamma P)  \right)d\tau
 -\int_A|\xi_r|^2(dP/dr) dS\right], \label{canE}
\end{equation}
where $\delta_{m,0})$ is the Kronecker $\delta.$
Here the first volume integral is taken over the planet
volume $V$ and the second surface integral is taken over 
the surface area $A.$ The contribution of the surface
term formally vanishes when the density at the surface
is zero. When the density at the surface is relatively small
as in our numerical model this term is found to give a negligible contribution.
When tidal forcing operates, the time rate of $E_c$
 gives the rate of energy uptake by the planet. After the encounter
tidal forcing ceases and $E_c$ is conserved in the absence of dissipation.
In fact, since some numerical diffusion is included, $E_c$ is observed to
slowly decay in some cases. This effect is, of course, more significant
if small scale disturbances are excited as in the case of models
with a significant solid core.

\subsubsection {The canonical Angular momentum}
Similarly, the canonical angular momentum is given by
\begin{equation}
J_c=-{\cal I} m\left[0.5m\int_V  \rho\left(   \xi_r^*v_r    +\xi_{\theta}^*v_{\theta} +\xi_{\phi}^*v_{\phi} 
+ 2\Omega (\xi_r\xi^*_{\phi}\sin\theta+  \xi_{\theta}\xi_{\phi}^* \cos\theta )\right)d\tau\right],
 \end{equation}
where ${\cal I}m$ indicates that the imaginary part is to be taken.
Here we recall that the state variables are complex.
This behaves in a manner  analogous to the canonical energy, but as applied
to the total angular momentum content of the planet.

\section{Analytical expressions for the energy and angular momentum transfer due to tides 
  associated with inertial modes}\label{BB}

Expressions for  the energy, $\Delta E_{m}$,
and angular momentum, $\Delta L_{m},$
transferred to the planet as a result
of tides associated
with azimuthal mode number $m$ acting during a parabolic encounter   were
derived in PI.
For completeness, we briefly review them below.
They are:
\begin{equation}
\Delta E_{m}={C_{m}\over (1+q)^{2}}\bar {\Omega}^{4}\sum_{k} \lbrace
\bar \sigma_{k}^{4}(4-\bar \sigma_{k}^{2})Q_{k}^{2}
I^{2}_{2,-m}(y) \rbrace {E_{*}\over \eta^{6}},  \label{eq 26}
\end{equation}
\begin{equation}
\Delta L_{2}={2C_{2}\over (1+q)^{2}}\bar {\Omega}^{3}\sum_{k} \lbrace
\bar \sigma_{k}^{3}(4-\bar \sigma_{k}^{2})Q_{k}^{2}
I^{2}_{2,-m}(y) \rbrace {L_{*}\over \eta^{5}},  \label{eq 27}
\end{equation}
where 
$\bar \sigma_{k}= \sigma_{k}/\Omega,$ with $\sigma_k$
being the mode eigenfrequency  and the sum is taken over all modes
with a particular value of $m.$
Contributions from $m=0$ and $m=2$ have been considered. In addition
$\bar \Omega = \Omega/\Omega_{p}$ and $y=\bar \Omega (\bar \sigma_{k}+m)$, 

The coefficients $C_{2}={3\over 16}$,
$C_{0}=3/4$ and the functions $I_{2,-m}(y)$ are determined by Fourier
transform of the tidal potential. The  latter quantities  are
described by  Press $\&$ Teukolsky 1977. They are exponentially small
for large values of the argument $y$.

The overlap integrals
$Q_{k}$ measure the  coupling of a particular mode with the tidal potential.
They are given by
\begin{equation}
Q_{k}=\left({\rho\over c_{s}^{2}}Fr^{2}P^{m}_{2}|W_{k}\right)/\sqrt{\bar N_{k}},
\label{eq 27a}
\end{equation}
where $W_k$ is the eigenfunction for mode $(k)$ , 
 $\bar N_{k}=N_{k}/\Omega^{2}$, $P^{m}_{2}$ is the associated
Legendre function  and $F(r)$ is a correction due to  the perturbation of the
gravitational potential of the planet. In the Cowling approximation  adopted here,
 $F=1$. The norm $N_{k}$ is given by
\begin{equation}N_{k}=\sigma^{2}_{k}(W_{k}| {\bmth {A}}
W_{k})+(W_{k}|{\bmth {C}} W_{k})\label{eq 16a}\end{equation}
 where the inner product is defined through
\begin{equation}
(W_{1}|W_{2})=\int_{V} dz \varpi d\varpi W^{*}_{1}W_{2},  \label{eq 7a}
\end{equation}
with $V$ being the planet volume and the operators ${\bf A}$ and
${\bf C}$ are defined below.

\subsection{Expressions for the energy and angular momentum transfer due to tides 
  in the anelastic impulsive limit}\label{BB1}
  We here explore the limit in which the characteristic time associated with the tidal  encounter
  $\Omega_p^{-1}$ is short compared to the rotation period.  In this limit we expect the
  excitation of inertial modes to be impulsive and thus the amount of energy
  and angular momentum transferred should  not depend on fine details  of the mode spectrum.
  This is because there is no time for inertial waves to propagate, reflect and indicate the presence
  of any possible standing waves,  critical  latitude phenomena  or  wave attractors. Later evolution would depend on this
  but the energy transferred would not. This suggests that a relatively simple
   expression for eg.  $\Delta E_m$ 
  should exist  which does not require detailed knowledge of the mode spectrum.
  We now show how to obtain this.

  The tidal response to an external potential with fixed frequency $\sigma$
and azimuthal mode number, $m,$ satisfies (IP) 
\begin{equation}
\sigma^{2}  {\bmth{A}}  W -\sigma {\bmth{B}}
 W -{\bmth{C}} W =
\sigma^{2}d{\rho \over c_{s}^{2}}
(\Psi_{ext,m}-W),  \label{eq g4b}
\end{equation}
\noindent where $d=4\Omega^{2}-\sigma^{2}$, and the operators
\begin{equation}
{\bmth {A}}=-{1\over \varpi}{\partial \over \partial \varpi}
\left (\varpi \rho{\partial\over \partial \varpi}\right )-{\partial \over \partial z}\left (\rho
{\partial \over \partial z}\right )+
{m^{2} \rho \over \varpi^{2}},  \label{eq 5b}
\end{equation}
\begin{equation}
{\bmth {B}} =-{2m\Omega \over \varpi}{\partial \rho \over \partial \varpi} \qquad  {\rm and}
\quad {\bmth {C}}=-4\Omega^{2}{\partial \over \partial z}\left(\rho {\partial
\over \partial z}\right).  \label{eq 6}
\end{equation}
 Adopting a zero density surface, the operators ${\bmth {A }}$,
${\bmth {B}}$ and ${\bmth {C}}$ are self-adjoint.
The eigenfrequencies, $\sigma=\sigma_k$ and eigenfunctions $W=W_{k}$ 
satisfy (\ref{eq g4b}) with $\Psi_{ext,m}$ set to zero.

In the anelastic approximation   $W$ is neglected on the right hand side
of (\ref{eq g4b})  (IPN) so that   the tidal response
then satisfies       
\begin{equation}
\sigma^{2}  {\bmth{A}}  W -\sigma {\bmth{B}}
 W -{\bmth{C}} W =
\sigma^{2}d{\rho \over c_{s}^{2}}
\Psi_{ext,m},  \label{eq 4b}
\end{equation}

To consider the impulsive limit, we restore the time dependence by replacing 
$\sigma$ by the operator $i\partial / \partial t$ and note that in the limit of interest
where the tidal interaction is fast, $\sigma$ may be considered to be large.
This suggests we seek a formal solution of (\ref{eq 4b}) as a series in $\sigma$ of the form
\begin{equation}
W=\sum_{j=0}^{\infty} {\cal F}_j\sigma^{2-j}\label{sumb}.
\end{equation}
Having obtained this we may  go on to find the impulsively induced velocity 
 using equation (\ref{eq p3}) without viscosity, which leads to 
\begin{equation}
\Delta {\bf v} = -\int^{\infty}_{-\infty}\nabla Wdt. \label{impb}
\end{equation}
Here we remark that although we have written infinite limits,
we have to suppose that there is a formal time scale separation
so the limits may be assumed large (and effectively infinite) in magnitude when
compared to $\Omega_p^{-1}$ but small when  compared to $\Omega^{-1}.$

We may then obtain the energy transferred by substituting this velocity into the expression
for the canonical energy
(\ref{canE}) while neglecting the pressure term $\propto \gamma^{-1}$ which is assumed negligible
in the anelastic approximation and  also the boundary term in that expression.
To evaluate (\ref{impb}) , recalling the operator interpretation of
$\sigma,$  the assumed time scale separation,  and the fact that 
the forcing potential may be assumed
to vanish before and after the encounter,  one may verify that 
 only the term  with $j=2$ in (\ref{sumb}) is needed to evaluate (\ref{impb}).
 This is readily found to be given by 
 \begin{equation}
   W= {\cal F}_2={\bf A}^{-1}\left( 4\Omega^2 -{\bf C}{\bf A}^{-1}
-{\bf B}{\bf A}^{-1}{\bf B}{\bf A}^{-1}\right)\cdot {\rho \over c_{s}^{2}}
\Psi_{ext,m}
   \end{equation}
Although this expression involves inversion
of the elliptic
 Laplacian like operator ${\bf A},$
which depends on  an inner boundary condition
that  could couple different orders  in a way 
 that has not been indicated above,
 this should be in principle straightforward. 
On following the procedure outlined above to calculate $\Delta E_m$
(the angular momentum transferred $,\Delta L_m,$ is formally  higher order
in the ratio $\Omega/\Omega_p$ so we  do not consider it further here)
one sees that it scales as $\Omega^4$ and that it  typically involves integral
operations on the perturbing potential over the planet volume.
Thus a small core is expected to have little influence on account of its small
volume irrespective of details of the spectrum.
We finally comment that although the above formal  argument required 
$\Omega/\Omega_p$ to be small, on physical grounds we would expect
similar conclusions to hold provided there is no time for
wave propagation with multiple reflections during the tidal encounter.

\section{ Analytic solutions and vanishing of overlap
integrals  for coreless models with fixed quadratic
gravitational potential}\label {Goodman models}
Goodman \& Lackner (2009)  considered special barotropic models
which are in hydrostatic equilibrium  under a fixed
quadratic gravitational potential $\Psi$ in the unperturbed state.
Thus
 \begin{equation}
   \frac{1}{\rho}\frac{dP}{dr} =  \frac{c^2_s}{\rho}\frac{d\rho}{dr}=-\frac{d\Psi}{dr},
   \end{equation}
with $\Psi = c_1+c_2r^2,$ with $c_1$ and $c_2$ being constant.

As the Cowling approximation of neglecting the perturbation
to the gravitational potential  is adopted, these models can be
treated in the same way as the standard  polytropic models we have considered
which are in hydrostatic equilibrium under a self-consistent gravitational
potential in the unperturbed state.

If the equation of state is regarded as being free, a  fixed quadratic
gravitational potential model can be constructed with any density distribution.
Thus  there will always be one corresponding to a polytropic distribution
which will have the same anelastic mode spectrum as this  depends only
on the density distribution (see  (\ref{eq 4b}) ). However,
such models have a special property in common with the uniform density
model that $(1/\varpi)\partial \Psi/\partial \varpi 
= (1/r)d\Psi/dr =c_2 \equiv \omega_0^2$ is constant. A consequence
of this is that there is an analytic 
solution that shows that no inertial modes are excited in response to
forcing by a quadrupole $(m=2)$ potential as was indeed pointed out
by Goodman \& Lackner (2009) in the case when the fully
compressible response governed by  equation (\ref{eq g4b}) is considered.
We here point out that this is also the case when the anelastic approximation
is used  and thus a spurious excitation  of inertial modes
does not occur in this situation as was suggested by Goodman \& Lackner (2009).

To solve the response problem for the quadratic potential models
with quadrupole forcing with $m=2,$ we set
$\Psi_{ext,m}= c_3\varpi^2$ where $c_3$ is constant (see equation(\ref{Tidpot}))
in the governing equation
\begin{equation}
\sigma^{2}  {\bmth{A}}  W -\sigma {\bmth{B}}
 W -{\bmth{C}} W =
\sigma^{2}d{\rho \over c_{s}^{2}}
(\Psi_{ext,m}-fW),  \label{eq h4b}
\end{equation}
where we have inserted the constant $f$ which allows us to pass between the
fully compressible case $f=1$ and the anelastic approximation $f=0.$
A solution can be found by setting $W=c_4\varpi^2$ in (\ref{eq h4b})
 where $c_4$ is a constant
to be determined. If this is done one readily finds that
\begin{equation}
\left(2\sigma(\sigma-2\Omega)\omega_0^2+f\sigma^2d\right)c_4=\sigma^2dc_3.
\end{equation}
 When $\sigma \ne 0$ for any value of $f$   or $\sigma \ne  2\Omega$  for  $f=0,$  this
straightforwardly  implies that
\begin{equation}
c_4=\frac{c_3\sigma(\sigma+2\Omega)}{(f\sigma^2+2\Omega f\sigma -2\omega^2_0)}.
\label{Goodmansol} \end{equation}
One may also verify that this solution applies to the
cases when  $\sigma=0$ and  $\sigma =2\Omega$ with $f=0$   as well,
there being zero response to zero forcing in  the former case.
In the latter case there is apparently a unique normal mode with $W \propto \varpi^2.$
However,  the form of the forcing means that this  also does  not produce
a resonant singularity in the response and therefore it is not associated with
any tidal energy or angular momentum transfer as is also 
implied by equations  (\ref{eq 26}) and (\ref{eq 27}).

In the fully compressible case, $f=1,$ (\ref{Goodmansol}) implies
singularities only  when $\sigma = -\Omega \pm\sqrt{2\omega_0^2+\Omega^2}$
which corresponds to the forward and backward propagating  $f$ modes.
This means that only these modes and no inertial modes would be excited
in a tidal encounter. In the anelastic case, there are no singularities at all
meaning no modes are excited. Also, as expected from
the discussion in IPN,  when $|\sigma|$ and  $\Omega$
are of the same order and small, the anelastic and fully compressible
solutions differ by the  order of $(\Omega/\omega_0)^2.$

It is of interest to connect the above solution to the 
energy and angular momentum exchanges during a tidal encounter  given by
(\ref{eq 26}) and ({\ref{eq 27}). From this we expect these 
exchanges to be zero, which in turn implies that for any inertial mode
that could contribute,
with $\sigma_k \ne 0,$ or $\sigma_k^2 \ne 4\Omega^2,$ the overlap
integral $Q_k =0.$ It is easy to verify that this is indeed the case
by writing down the anelastic normal mode equation
\begin{equation}
{\cal L}(W_k)\equiv \sigma_k^{2}  {\bmth{A}}  W _k -\sigma_k {\bmth{B}}
 W_k-{\bmth{C}} W_k =0  \label{eq i4b}
\end{equation}
and considering the inner product
\begin{equation}
\left(\varpi^2 | {\cal L}(W_k)\right)=0.
\end{equation}
From this one obtains
\begin{equation}
\left(\frac{\rho\varpi^2\omega_0^2}{ c_s^2} | W_k\right)=0
\end{equation}
which confirms that indeed the overlap integrals are all zero
when $\omega_0^2$ is constant.
For the  global modes  that could potentially
have significant overlap integrals, adopting appropriate relative error
estimates,  this result  is well
represented using spectral approaches of the kind considered
in Papaloizou \& Pringle (1981) and IP as well as the finite difference
approach considered in this paper.

 We emphasise again that the vanishing of the overlap integrals
for the quadratic potential models depends on the
constancy of $\omega_0^2.$
In realistic planetary or stellar models, 
that are in hydrostatic equilibrium under their self consistent gravitational potential,
the amount of variation of
$\omega_0^2$ is measured by the central condensation
or the ratio of central to mean density which differs significantly
from unity  even for a standard  polytrope of index $n=1$ as considered
in this paper. Accordingly there is no reason to suppose
that the overlap integrals should be zero in that case.}

\end{appendix}

\bsp

\label{lastpage}

\end{document}